\documentclass[journal]{IEEEtran}

\usepackage{fontenc}
\usepackage{lmodern}
\usepackage{microtype}

\usepackage[noadjust]{cite}
\usepackage{graphicx}
\usepackage[caption=false,format=hang]{subfig}
\usepackage{physics}
\usepackage{amssymb}
\usepackage{hyperref}
\usepackage[capitalise,compress,nameinlink]{cleveref}
\usepackage[abbreviations,british]{foreign}
\usepackage{interval}
\usepackage[separate-uncertainty=true]{siunitx}
\usepackage{booktabs}
\usepackage{overpic}
\usepackage{xcolor}

\renewcommand{\abs}[1]{\big\lvert#1\big\rvert}
\renewcommand{\norm}[1]{{\big\lVert#1\big\rVert}_{2}}

\newcommand{\almost}[1]{{\thicksim}#1}
\newcommand{\conj}[1]{#1^{\ast}}
\newcommand{\convolution}[2]{(#1 \circledcirc{} #2)}
\newcommand{\diag}[1]{\text{diag}\{#1\}} 
\newcommand{\edges}{\mathcal{E}}
\newcommand{\faces}{\mathcal{F}}
\newcommand{\graph}{\mathcal{G}}
\newcommand{\hermitian}[1]{#1^{\dagger}}
\newcommand{\hilbert}[1]{L^{2}(#1)} 
\newcommand{\imax}{i_{\text{max}}}
\newcommand{\integrateManifold}[1]{\displaystyle\int\limits_{\mathcal{M}} \dd{#1}}
\newcommand{\integrateRegion}[1]{\displaystyle\int\limits_{R} #1}
\newcommand{\laplace}[1]{\Delta_{\mathcal{#1}}}
\newcommand{\manifold}{\mathcal{M}}
\newcommand{\mesh}[1]{#1(x)}
\newcommand{\meshSum}{\displaystyle\sum\limits_{i}}
\newcommand{\meshVolume}{\dd{x}}
\newcommand{\meshYSum}{\displaystyle\sum\limits_{j}}
\newcommand{\meshY}[1]{#1(y)}
\newcommand{\realPosParam}{\mathbb{R}_{\ast}^{+}}
\newcommand{\set}[1]{\{#1\}} 
\newcommand{\slepian}[2][]{{#2}_{p#1}}
\newcommand{\slepianSum}[1][]{\displaystyle\sum\limits_{p#1}}
\newcommand{\snr}[1]{\text{SNR}(#1)} 
\newcommand{\translation}[1]{\mathcal{T}_{#1}}
\newcommand{\twoSphere}{\mathbb{S}^{2}}
\newcommand{\variance}[1]{{\big[\Delta#1\big]}^{2}}
\newcommand{\vertices}{\mathcal{V}}
\newcommand{\waveletSum}{\displaystyle\sum\limits_{j=J_{0}}^{J}}

\intervalconfig{soft open fences}

\graphicspath{{figures/}}

\begin{document}

\title{Slepian Scale-Discretised Wavelets on Manifolds}

\author{
	\IEEEauthorblockN{
		Patrick J. Roddy\IEEEauthorrefmark{1} and
		Jason D. McEwen\IEEEauthorrefmark{1}
	}

	\IEEEauthorblockA{
		\IEEEauthorrefmark{1}
		University College London (UCL), Gower Street, London, WC1E 6BT, UK
		\newline
		\footnotesize \texttt{patrick.roddy@ucl.ac.uk}, \texttt{jason.mcewen@ucl.ac.uk}
	}
}

\maketitle

\begin{abstract}
    Inspired by recent interest in geometric deep learning, this work generalises the recently developed Slepian scale-discretised wavelets on the sphere to Riemannian manifolds.
    Through the sifting convolution, one may define translations and, thus, convolutions on manifolds --- which are otherwise not well-defined in general.
    Slepian wavelets are constructed on a region of a manifold and are therefore suited to problems where data only exists in a particular region.
    The Slepian functions, on which Slepian wavelets are built, are the basis functions of the Slepian spatial-spectral concentration problem on the manifold.
    A tiling of the Slepian harmonic line with smoothly decreasing generating functions defines the scale-discretised wavelets; allowing one to probe spatially localised, scale-dependent features of a signal.
    By discretising manifolds as graphs, the Slepian functions and wavelets of a triangular mesh are presented.
    Through a wavelet transform, the wavelet coefficients of a field defined on the mesh are found and used in a straightforward thresholding denoising scheme.
\end{abstract}


\IEEEpeerreviewmaketitle{}

\section{Introduction}

Many fields measure data that are intrinsically non-Euclidean in structure and are better modelled by manifolds or graphs.
One manifold, in particular, which is commonplace in science and engineering is the sphere; such as in: cosmology~\cite{Bennett1996}, geophysics~\cite{Simons2006}, planetary science~\cite{Turcotte1981}, computer graphics~\cite{Ramamoorthi2004}, and signal processing~\cite{Roddy2021,Roddy2022}.
Often data are not observed in some regions of the manifold, and hence methods which work over the whole manifold may not be appropriate.
One such example, in the spherical setting, is in CMB analyses where foreground microwave emissions dominate the region around the Galactic plane and are often removed~\cite{Mortlock2002}.
Wavelets are a typical approach to deal with problems of this form, which allow the probing of spatially localised, scale-dependent features of the signal.
Contamination of the wavelet coefficients at the boundaries of the region, however, still presents a problem.
To overcome this, in~\cite{Roddy2022}, Slepian wavelets are constructed within a region of the sphere.
Here, in this work, Slepian wavelets are generalised to the manifold setting following an analogous construction.

In data analyses, one often desires to extract non-trivial patterns in the data.
By projecting the data onto an appropriate basis, patterns that were not considered before may present themselves.
If one seeks to find oscillatory features, then a Fourier basis is often most appropriate, \ie{} the spherical harmonics in the spherical setting.
Wavelets, however, simultaneously extract contributions of scale-dependent features in both space and scale.
Wavelets on the sphere (\cf{} manifold) have been applied in fields such as: astrophysics and cosmology~\cite{Pen1999,Barreiro2001,Rocha2004,McEwen2004}, planetary science~\cite{Audet2011,Audet2014}, and geophysics~\cite{Loris2010,Simons2011,Simons2011b}.
Scale-discretised wavelets (on the sphere)~\cite{Wiaux2008,McEwen2018,Leistedt2013,McEwen2013,McEwen2015} lean on a tiling of the harmonic line to produce an exact wavelet transform in both the continuous and discrete settings.

A new research direction, geometric deep learning, has evolved from an increasing interest in non-Euclidean data geometries; where convolutional networks are generalised to graph and manifold structured data (\eg{}~\cite{Bronstein2017,Perlmutter2020}).
In general, geometric learning problems are split into two categories: characterising the structure of the data and analysing functions defined on a non-Euclidean domain.
Manifold learning models (\eg{}~\cite{Tenenbaum2000,Coifman2006b,VanDerMaaten2008}) are a set of unsupervised algorithms which capture the intrinsic structure in the data through data-driven geometries.
Moreover, signals on manifolds are increasingly prevalent in areas such as shape matching and computer graphics.
Thus, many works are emerging to generalise spectral and signal processing methods to manifolds~\cite{Coifman2006} and graphs~\cite{Shuman2013}.
In these settings, functions are supported on the manifold/the vertices of the graph, and the Fourier harmonics are the eigenfunctions of the Laplace-Beltrami operator/eigenvectors of the graph Laplacian.

Graphs are a fundamental data structure, which are useful in describing geometric structures of data in domains such as: social~\cite{Nettleton2013}, transportation~\cite{Mohan2014}, sensor~\cite{Kenniche2010}, and neuronal~\cite{Tang2012} networks.
Each edge in a graph is associated with a weight, which is often a measure of the similarity between two connected vertices in a graph.
The physics of the problem or the data at hand dictate which vertices are connected and their respective weights, \ie{} the weight may not necessarily be the physical distance between two vertices.
A graph signal is a finite collection of samples with some value at each vertex of a graph.
In computer graphics, working with intrinsically geometric shapes is commonplace.
In this field, three-dimensional shapes are often modelled as Riemannian manifolds and discretised as triangular meshes.

It is well-known that functions cannot have finite support in the spatial and spectral domains simultaneously~\cite{Slepian1961,Slepian1983}.
In the 1960s, Slepian, Landau and Pollak solved the fundamental problem of finding and representing signals which are optimally concentrated in both domains~\cite{Slepian1961,Landau1961,Landau1962}.
The Slepian spatial-spectral concentration problem (or the Slepian concentration problem for short) results in the families of functions that are optimally concentrated in the spatial (spectral) domain, and exactly limited in the spectral (spatial) domain.
The Slepian concentration problem was initially formed in the Euclidean domain; however, it has since been generalised to other geometries, such as: the sphere~\cite{Simons2006,Wieczorek2005,Albertella1999,Cohen1989,Meaney1984,Daubechies1988}, and graphs~\cite{VanDeVille2017,VanDeVille2017a,Bolton2018}.
The so-called Slepian functions arise in many fields within science and engineering, such as: medical imaging~\cite{Jackson1991}, signal processing~\cite{Mathews1985,Thomson1982}, and geophysics~\cite{Thomson1976,Simons2006a,Simons2011}.
Notably, these functions have been used in interpolation~\cite{Moore2004,Shkolnisky2006}, extrapolation~\cite{Xu1983}, inverse problems~\cite{Villiers2001,Abdelmoula2015}, and solving partial differential equations~\cite{Boyd2003,Chen2005}.

Due to recent interest in geometric deep learning, many fields are extending existing methods developed in the Euclidean domain to manifolds or graphs, \ie{}~\cite{Perlmutter2020}.
This work generalises Slepian scale-discretised wavelets on the sphere~\cite{Roddy2022} to Riemannian manifolds.
These wavelets constitute a basis designed for incomplete data on manifolds or graphs.
Here, the scale-discretised wavelet construction on the sphere~\cite{Wiaux2008,McEwen2018,Leistedt2013,McEwen2013,McEwen2015} is extended to the Slepian domain; however, the Slepian functions are now constructed on a region of a general manifold (rather than strictly the sphere).
Scale-discretised wavelets exhibit an explicit inversion formula, constitute a tight frame, and have excellent localisation properties in both spectral and spatial domains.
The wavelets themselves are constructed through a tiling of the Slepian harmonic line --- built from the eigenfunctions of the Slepian concentration problem on the manifold.
The wavelet transform follows through a generalisation of the sifting convolution~\cite{Roddy2021}, which allows one to perform convolutions on (incomplete) manifolds.
The wavelet construction is analogous to the spherical case, but with the eigenfunctions of the Laplace-Beltrami operator/eigenvectors of the graph Laplacian, rather than the spherical harmonics.

Slepian scale-discretised wavelets on the sphere were introduced before in~\cite{Roddy2022}.
This work generalises these wavelets to Riemannian manifolds; however, wavelets in the Slepian domain have been considered before.
Multiresolution analyses have been performed with prolate spheroidal wave functions~\cite{Walter2004} in the Euclidean setting.
In spherical settings, spatially localised spherical harmonic transforms are sometimes used (\eg{}~\cite{Simons1997,Wieczorek2005,Khalid2013,Khalid2013a}).
Other times, the Slepian functions on the sphere are used directly~\cite{Simons2009}.
A variety of other approaches include: established regularisation techniques based on a known singular value decomposition~\cite{Michel2017}, and wavelet-like representations without explicit inverse transforms~\cite{Simons2011}.
Extending wavelets and wavelet techniques to manifold and graph domains has been extensively reviewed (\eg{}~\cite{Dahmen1999,Coifman2006a}).

The remainder of this article is as follows.
\cref{sec:mathematical_background_problem_formulation} presents some mathematical preliminaries of Riemannian manifolds, their representations as graphs, and triangular meshes.
The Laplace-Beltrami operator and the graph/mesh Laplacian are also introduced.
A brief review of Slepian scale-discretised wavelets on the sphere~\cite{Roddy2022} follows, which this work builds upon.
In \cref{sec:working_with_manifolds} the Slepian concentration problem is adapted to the manifold setting, which results in the basis functions of a region of the manifold, \ie{} the Slepian functions.
A review of the sifting convolution~\cite{Roddy2021} follows, which allows one to perform convolutions on the (incomplete) manifold --- which are typically not well-defined.
The Slepian wavelet theory is developed in \cref{sec:slepian_wavelets}.
Initially, the sifting convolution is expressed in the Slepian domain, which is a product in the Slepian harmonic space.
The scale-discretised framework is then introduced, followed by the generating functions which define the wavelets themselves.
The section concludes with some properties of the wavelets.
A numerical illustration is presented in \cref{sec:numerical_illustration}, in which the Slepian functions of the head region of a triangular mesh of Homer Simpson are computed.
A field is then placed on the mesh, and the resulting wavelets and wavelet coefficients are found through a wavelet transform.
A straightforward hard-thresholding denoising scheme is developed and performed on the Homer mesh, along with some other meshes.
A boost in the signal-to-noise ratio is observed, showcasing a possible use of these wavelets.
Lastly, some concluding remarks are given in \cref{sec:conclusions}.

\section{Mathematical Background and Problem Formulation}\label{sec:mathematical_background_problem_formulation}

Some mathematical preliminaries are introduced in \cref{sec:mathematical_preliminaries} with a recap of Riemannian manifolds, their representation as graphs, and triangular meshes (a popular choice of graph).
This includes a review of the Laplace-Beltrami operator (on manifolds) and the graph/mesh Laplacian.
This work extends Slepian scale-discretised wavelets on the sphere to manifolds, a brief review of which is given in \cref{sec:problem_formulation}.

\subsection{Mathematical Preliminaries}\label{sec:mathematical_preliminaries}

\subsubsection{Riemannian Manifolds}

In short, a manifold is a locally Euclidean space.
For example, at a point on the surface of the Earth (\ie{} \(\twoSphere{}\)) the Earth's surface appears planar --- leading to the so-called flat Earth theory.
More formally, a \(d\)-dimensional manifold \(\manifold{}\) is a topological space, where each point \(x\) exists in a neighbourhood that is topologically homeomorphic to a \(d\)-dimensional Euclidean space --- called the tangent space.
Let \(\manifold{}\) denote a compact, smooth, connected \(d\)-dimensional Riemannian manifold without boundary contained in \(\mathbb{R}^{n}\).
The Hilbert space \(\hilbert{\manifold}\) is formed by the set of functions \(f : \manifold \to \mathbb{R}\) that are square-integrable with respect to the Riemannian volume \(\meshVolume{}\).
The geodesic distance between two points is denoted \(r(x,y)\), and the Laplace-Beltrami operator on \(\manifold{}\) is denoted \(\laplace{M}\).

Let \(f \in \hilbert{\manifold}\) be a real-valued function defined on a Riemannian manifold \(\manifold{}\).
The Laplace-Beltrami operator is
\begin{equation}
	\laplace{M} f
	= \div(\grad{f}),
\end{equation}
where \(\div{}\) and \(\grad{}\) are the divergence and gradient operators on the manifold \(\manifold{}\) respectively.
The Laplacian eigenproblem is defined as
\begin{equation}
	\laplace{M} f
	= -\mu f,
\end{equation}
which admits an orthogonal eigensystem due to the semi-positive definiteness of the Laplace-Beltrami operator.
This eigensystem forms the square-integrable basis of the manifold \(\manifold{}\), with real eigenvalues \(0 \leq \mu_{1} \leq \mu_{2} \leq \ldots < \mu_{\imax} < \infty{}\), and eigenfunctions \(\zeta_{i}\). 
A function \(f \in \hilbert{\manifold}\) can therefore be decomposed into this basis by
\begin{equation}
	\mesh{f}
	= \sum\limits_{i=1}^{\imax} f_{i} \mesh{\zeta_{i}},
\end{equation}
where \(f_{i}\) are the Fourier coefficients given by the usual projection onto the basis functions \(f_{i} = \braket*{f}{\zeta_{i}}\).
In practice, one may discretise the manifolds and represent them as graphs.
A review of graphs is presented in \cref{sec:graphs}.

\subsubsection{Graphs}\label{sec:graphs}

For simplicity, consider a weighted undirected graph \(\graph = (\vertices, \edges)\), consisting of a set of \(n\) vertices \(\vertices \in \set{1,2,\ldots,n}\), and the set of edges \(\edges \subseteq \vertices \times \vertices{}\).
A weight \(a_{i} > 0\) is associated with each vertex \(i \in \vertices{}\), and a weight \(w_{i j} \geq 0\) with each edge \((i,j) \in \edges{}\).
The Hilbert spaces \(\hilbert{\vertices}\) and \(\hilbert{\edges}\) are formed by the sets of functions \(f: \vertices \to \mathbb{R}\) and \(F: \edges \to \mathbb{R}\) respectively.
The graph gradient operator \(\grad{}: \hilbert{\vertices} \to \hilbert{\edges}\) maps functions defined on the vertices to edges
\begin{equation}\label{eq:graph_grad}
	{(\grad{f})}_{i j}
	= f_{i} - f_{j}.
\end{equation}
The graph divergence operator \(\div: \hilbert{\edges} \to \hilbert{\vertices}\) reverses this mapping
\begin{equation}\label{eq:graph_div}
	{(\div F)}_{i}
	= \frac{1}{a_{i}} \sum\limits_{j : (i,j) \in \edges} w_{i j} F_{i j}.
\end{equation}
The graph Laplacian is an operator \(\laplace{G} : \hilbert{\vertices} \to \hilbert{\vertices}\) defined as
\begin{equation}\label{eq:graph_laplace}
	\laplace{G} f
	= \div(\grad{f}),
\end{equation}
which converges to the Laplace-Beltrami operator as the number of samples goes to infinity~\cite{Belkin2007}.

By substituting the definitions of the gradient \cref{eq:graph_grad} and divergence operators \cref{eq:graph_div} on the graph into \cref{eq:graph_laplace}, one finds
\begin{equation}\label{eq:graph_laplace_indices}
	{(\laplace{G} f)}_{i}
	= \frac{1}{a_{i}} \sum\limits_{(i,j) \in \edges} w_{i j} (f_{i} - f_{j}).
\end{equation}
Intuitively, this captures the geometric interpretation of the Laplacian --- as the difference between the local average of a function around a point and the value of the function itself at that point.
\cref{eq:graph_laplace_indices} can be rewritten in matrix form as
\begin{equation}\label{eq:graph_laplace_matrix}
	\vb*{\Delta} \vb*{f}
	= \vb*{A}^{-1} (\vb*{K} - \vb*{W}) \vb*{f},
\end{equation}
where \(\vb*{W} = (w_{i j})\) are the edge weights,
\begin{equation}
    \vb*{A}
    = \diag{a_{1},a_{2},\ldots,a_{n}}
\end{equation}
are the vertex weights,
\begin{equation}
    \vb*{K}
    = \diag{\sum_{j: j \neq i} w_{i j}}
\end{equation}
is the degree matrix, and the function \(f \in \hilbert{\vertices}\) is written as a column vector \(\vb*{f}\).
The non-normalised graph Laplacian refers to setting \(\vb*{A} = \vb*{I}\) in \cref{eq:graph_laplace_matrix}; other choices exist, such as the random walk Laplacian which occurs when \(\vb*{A} = \vb*{K}\)~\cite{VonLuxburg2007}.

\subsubsection{Triangular Meshes}

In computer graphics applications, three-dimensional shapes are often modelled by locally two-dimensional manifolds.
A manifold is sampled at a set of \(n\) points, and a graph is then constructed on these points (vertices) where the edges represent the local connectivity of the manifold.
A set of edge weights is then computed, for example, Gaussian weights
\begin{equation}
	w_{i j}
	= \exp(-\frac{\ell_{i j}^{2}}{2\sigma^{2}}),
\end{equation}
where \(\ell_{i j}\) represents the length of the edge between the \(i\) and \(j\) vertices.
However, this discretisation does not fully capture the geometry of the underlying continuous manifold; hence, the graph Laplacian would typically not converge to the Laplace-Beltrami operator as the sampling density increases~\cite{Bronstein2017,Wardetzky2007}.
By introducing another structure of faces \(\faces \in \vertices \times \vertices \times \vertices{}\), one can construct as geometrically consistent discretisation as possible.
The continuous manifold is represented by the polyhedral surface of connected triangles.

\begin{figure}
	\centering
	\begin{overpic}
		[width=.62\columnwidth]{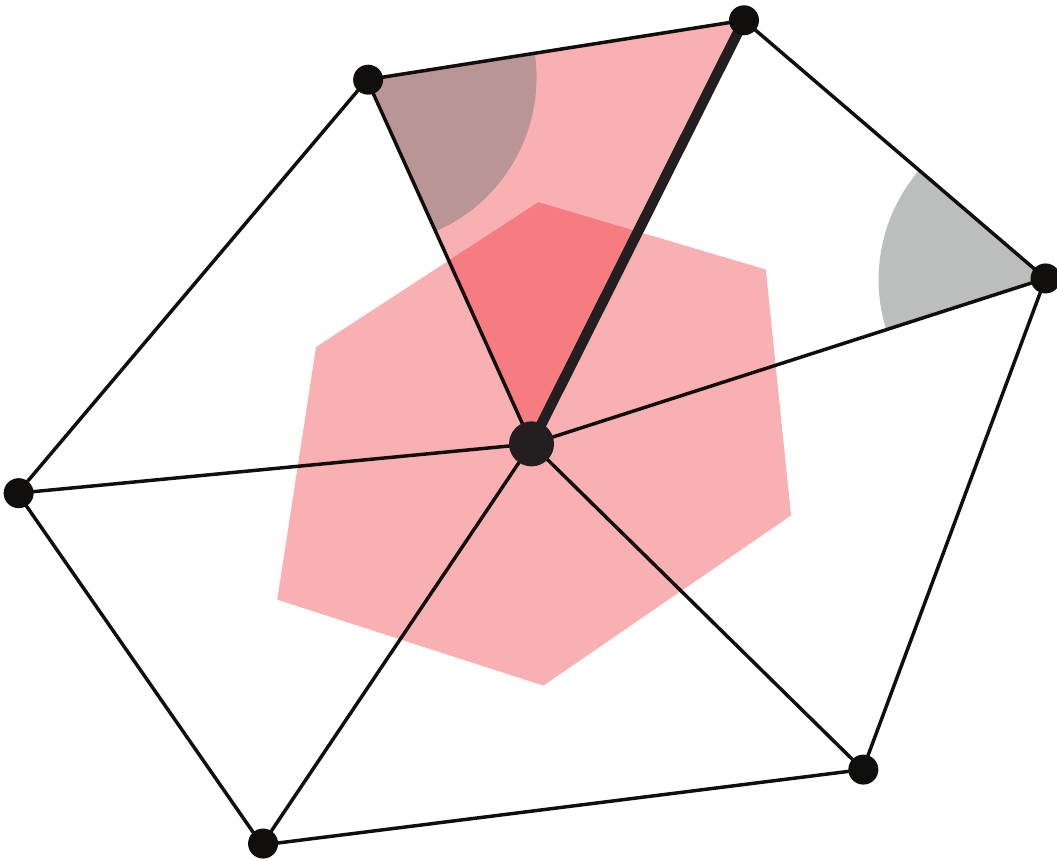}
		\small
		\put(101,53){\(h\)}
		\put(48,32){\(i\)}
		\put(69,83.5){\(j\)}
		\put(28,72.5){\(k\)}
		\put(38,69){\(\alpha_{i j}\)}
		\put(85,56){\(\beta_{i j}\)}
		\put(35.5,44.5){\color{red}\(a_{i}\)}
		\put(50,67){\color{red}\(a_{i j k}\)}
		\put(65,60){\(\ell_{i j}\)}
	\end{overpic}
	\caption{
		The triangular mesh discretisation of a two-dimensional manifold.
		A triangular face is highlighted in red which connects the \(i\), \(j\) and \(k\) vertices, with the corresponding area represented as \(a_{i j k}\).
		Similarly, the vertex weight \(a_{i}\) is the red hexagonal shape which is effectively one-third of the area of the surrounding faces, \ie{} \cref{eq:vertex_weight}.
		The \(i\) and \(j\) vertices are connected by an edge of length \(\ell_{i j}\), and the corresponding interior angles at the vertices \(k\) and \(h\) are given by \(\alpha_{i j}\) and \(\beta_{i j}\) respectively.
		These interior angles (in grey) appear in the expression for the cotangent Laplacian, \cf{} \cref{eq:cotangent_laplace}.
	}\label{fig:mesh_laplace}
\end{figure}
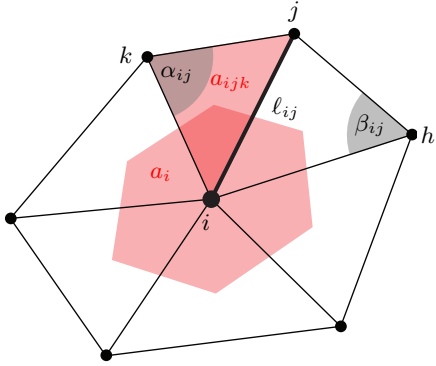

For triangular meshes, the most straightforward discretisation of the Riemannian metric is performed by assigning each edge a length \(\ell_{i j}>0\).
Consider such a discretisation shown in \cref{fig:mesh_laplace}.
The mesh Laplacian is then found by first substituting the edge weights into \cref{eq:graph_laplace_indices}
\begin{equation}\label{eq:mesh_laplace_weights}
	w_{i j}
	= \frac{-\ell_{i j}^{2} + \ell_{j k}^{2} + \ell_{i k}^{2}}{8 a_{i j k}}
	+ \frac{-\ell_{i j}^{2} + \ell_{j h}^{2} + \ell_{i h}^{2}}{8 a_{i j h}},
\end{equation}
where \(a_{i j k}\) represents the area of the triangular face connecting the \(i\), \(j\) and \(k\) vertices~\cite{Bronstein2017,Pinkall1993}.
The corresponding vertex weights of the mesh Laplacian are
\begin{equation}\label{eq:vertex_weight}
	a_{i}
	= \frac{1}{3} \sum\limits_{j k : (i,j,k) \in \faces} a_{i j k},
\end{equation}
which results in the red hexagonal region in \cref{fig:mesh_laplace}.
Such a construction can be shown to converge to the continuous Laplacian of the underlying manifold under some specific conditions~\cite{Wardetzky2008}.

\begin{figure}
	\centering
	\subfloat[\(\mesh{\zeta_{3}}\)]
	{\includegraphics[trim={156 8 21 6},clip,width=.25\columnwidth]{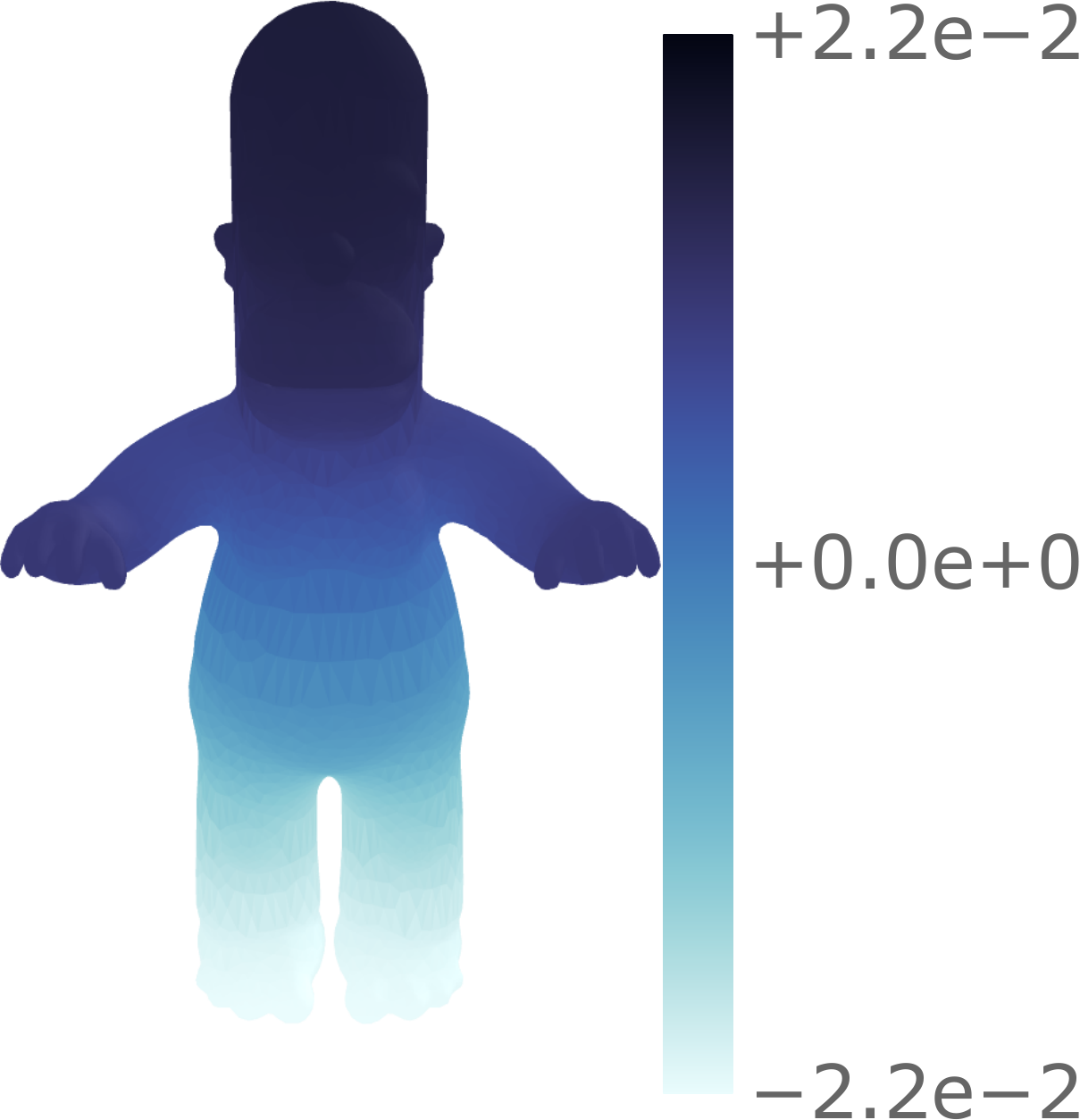}} 
	\hfill
	\subfloat[\(\mesh{\zeta_{4}}\)]
	{\includegraphics[trim={156 8 21 6},clip,width=.25\columnwidth]{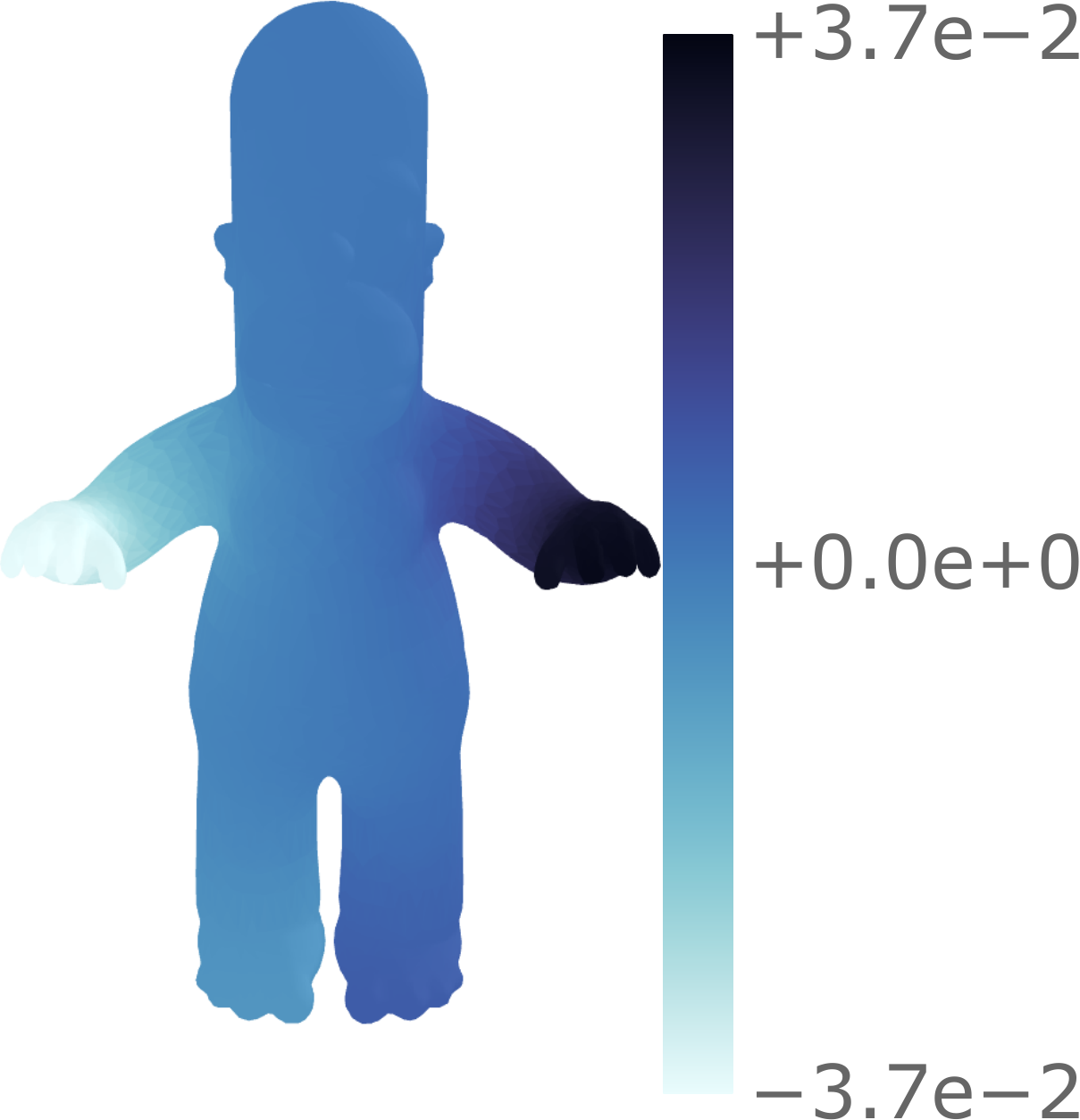}} 
	\hfill
	\subfloat[\(\mesh{\zeta_{5}}\)]
	{\includegraphics[trim={156 8 21 6},clip,width=.25\columnwidth]{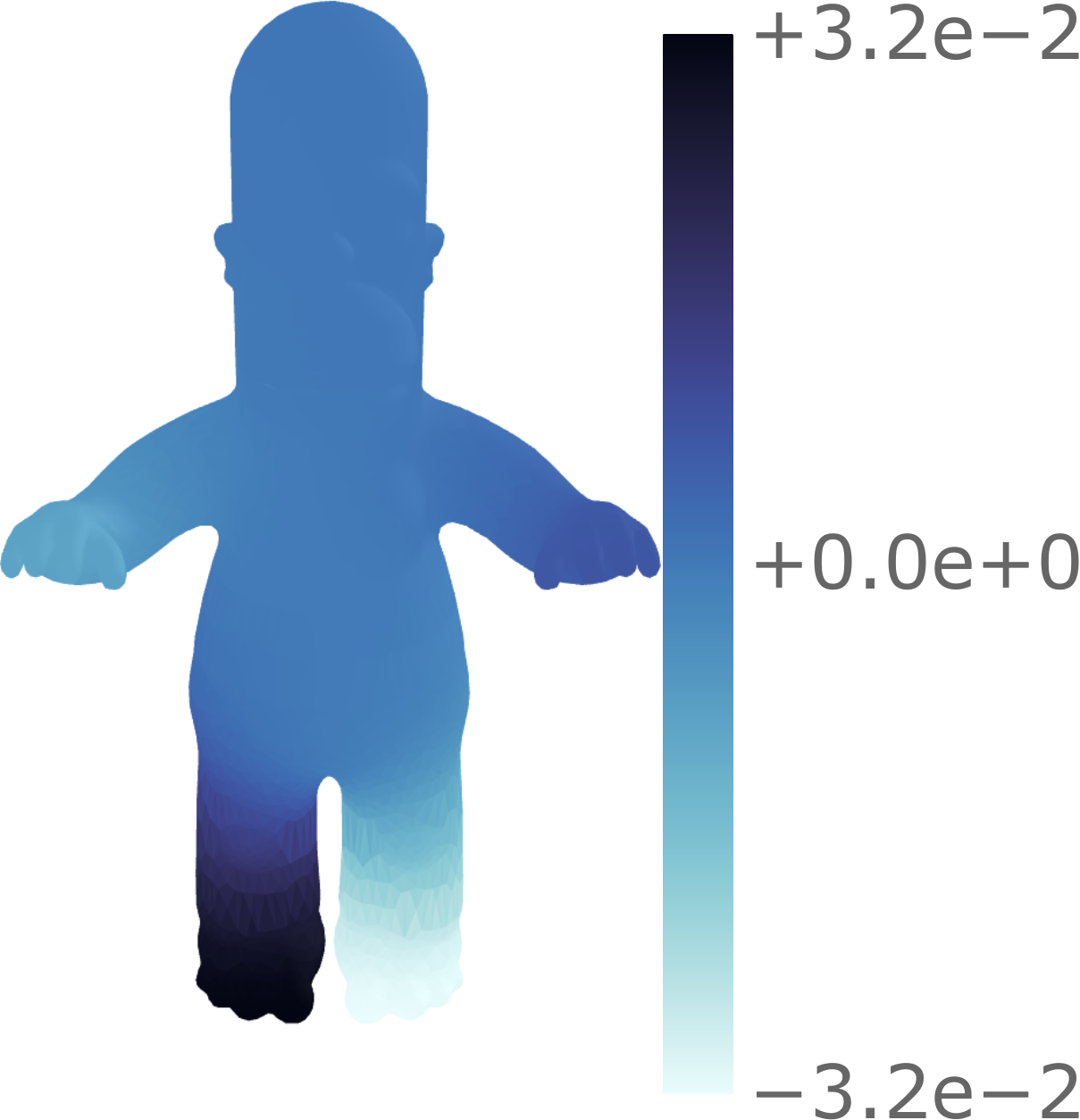}} 
	\hfill
	\subfloat[\(\mesh{\zeta_{6}}\)]
	{\includegraphics[trim={156 8 21 6},clip,width=.25\columnwidth]{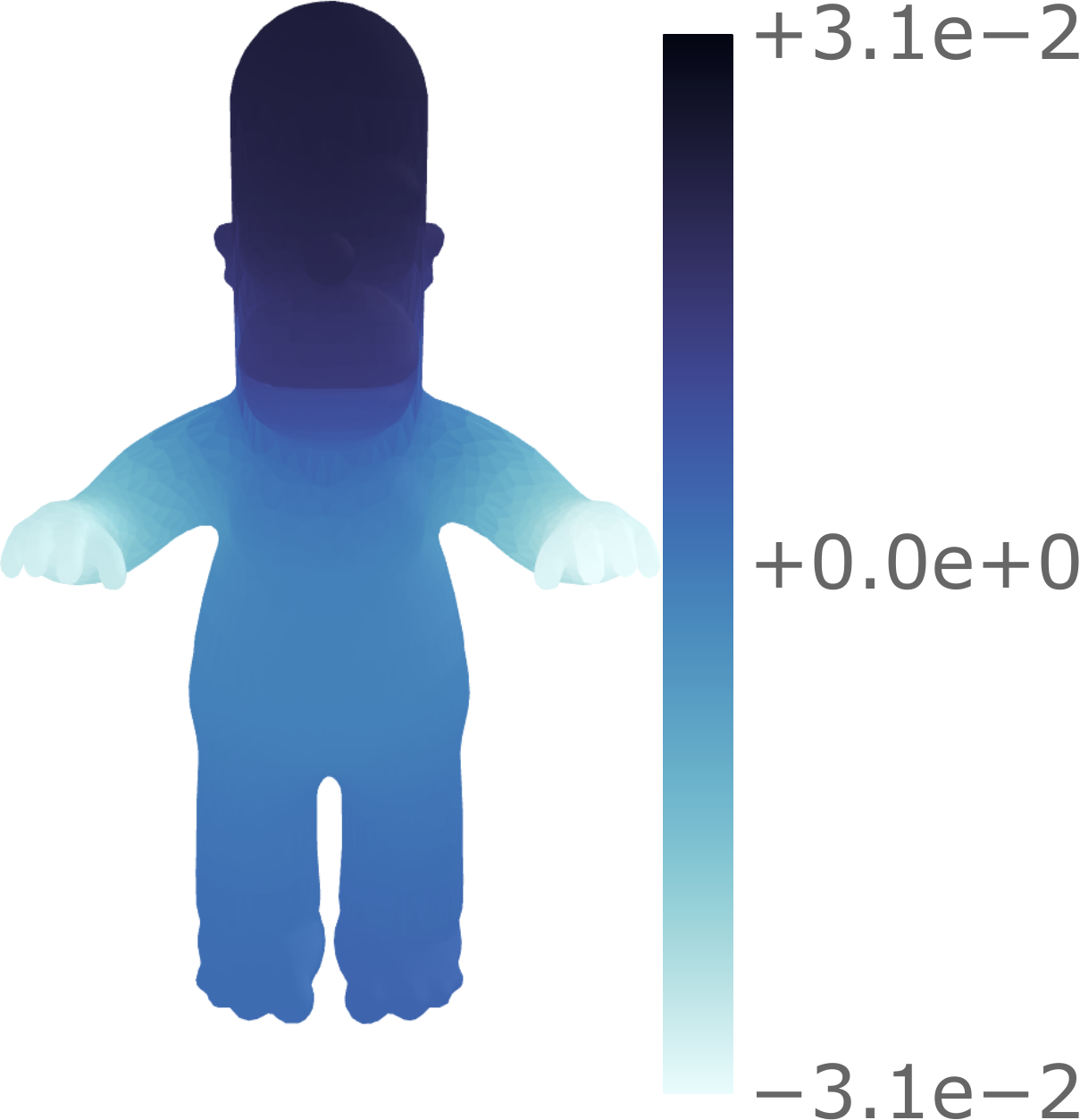}} 
	\newline
	\subfloat[\(\mesh{\zeta_{7}}\)]
	{\includegraphics[trim={156 8 21 6},clip,width=.25\columnwidth]{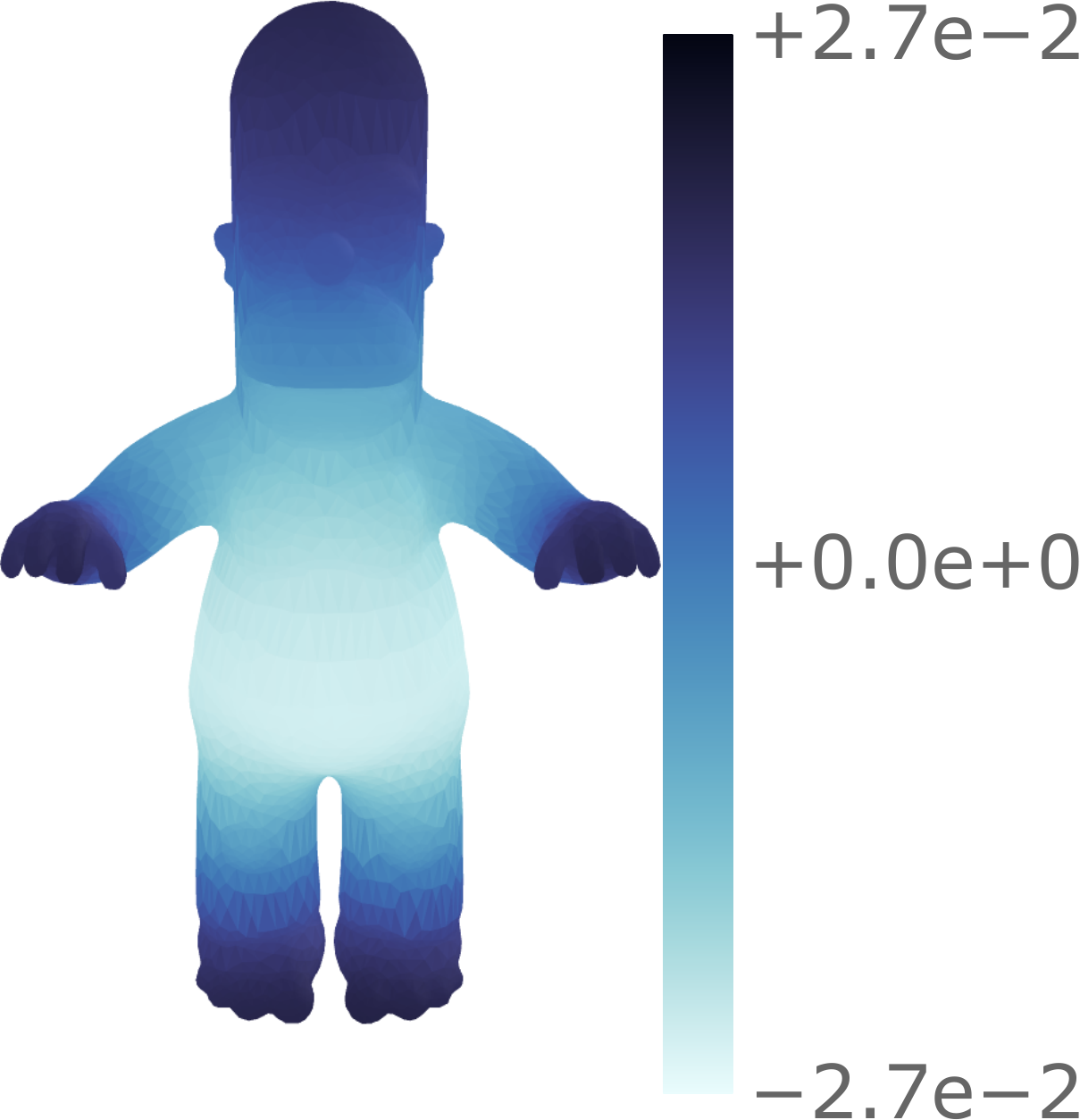}} 
	\hfill
	\subfloat[\(\mesh{\zeta_{8}}\)]
	{\includegraphics[trim={156 8 21 6},clip,width=.25\columnwidth]{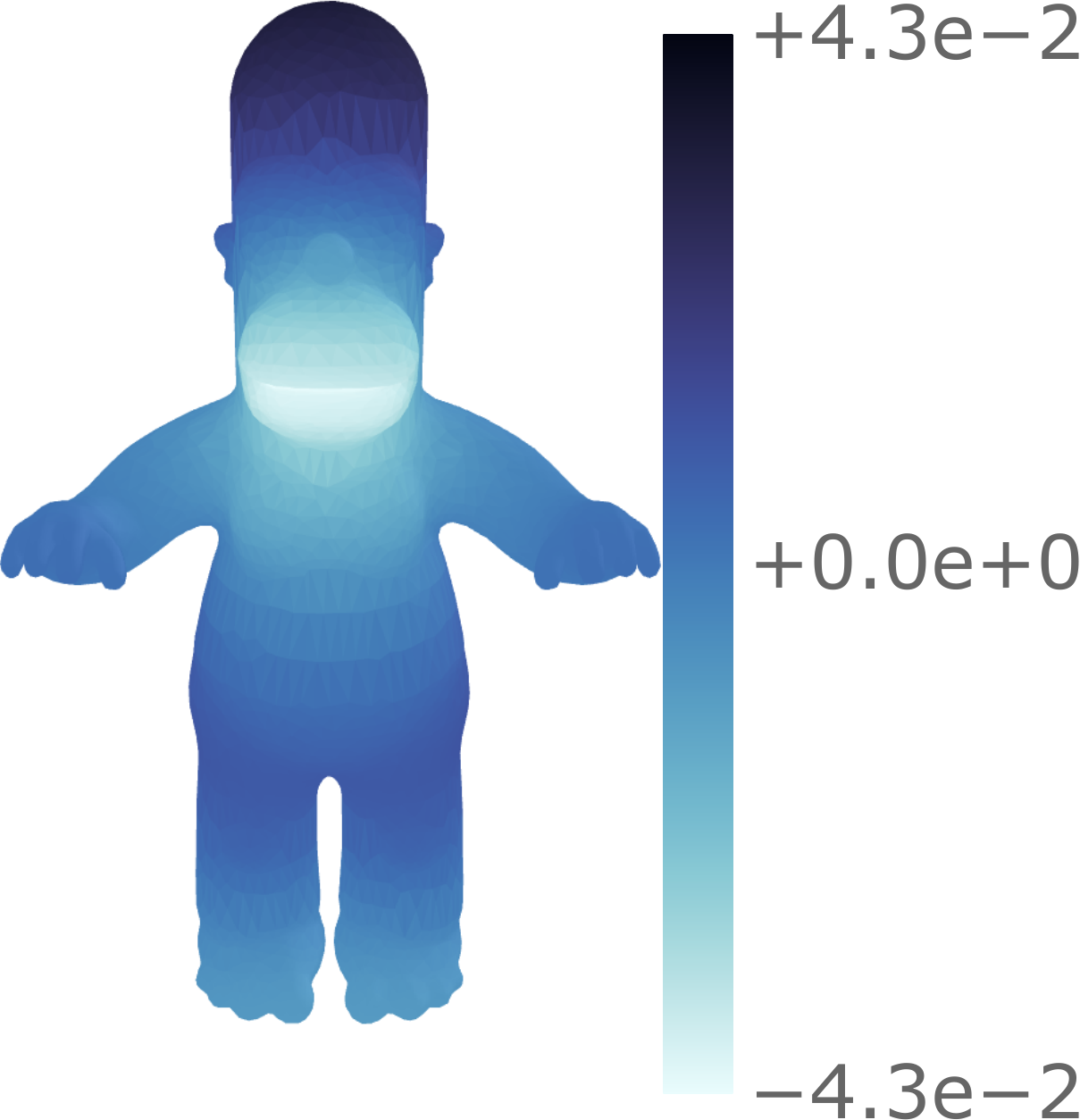}} 
	\subfloat[\(\mesh{\zeta_{9}}\)]
	{\includegraphics[trim={156 8 21 6},clip,width=.25\columnwidth]{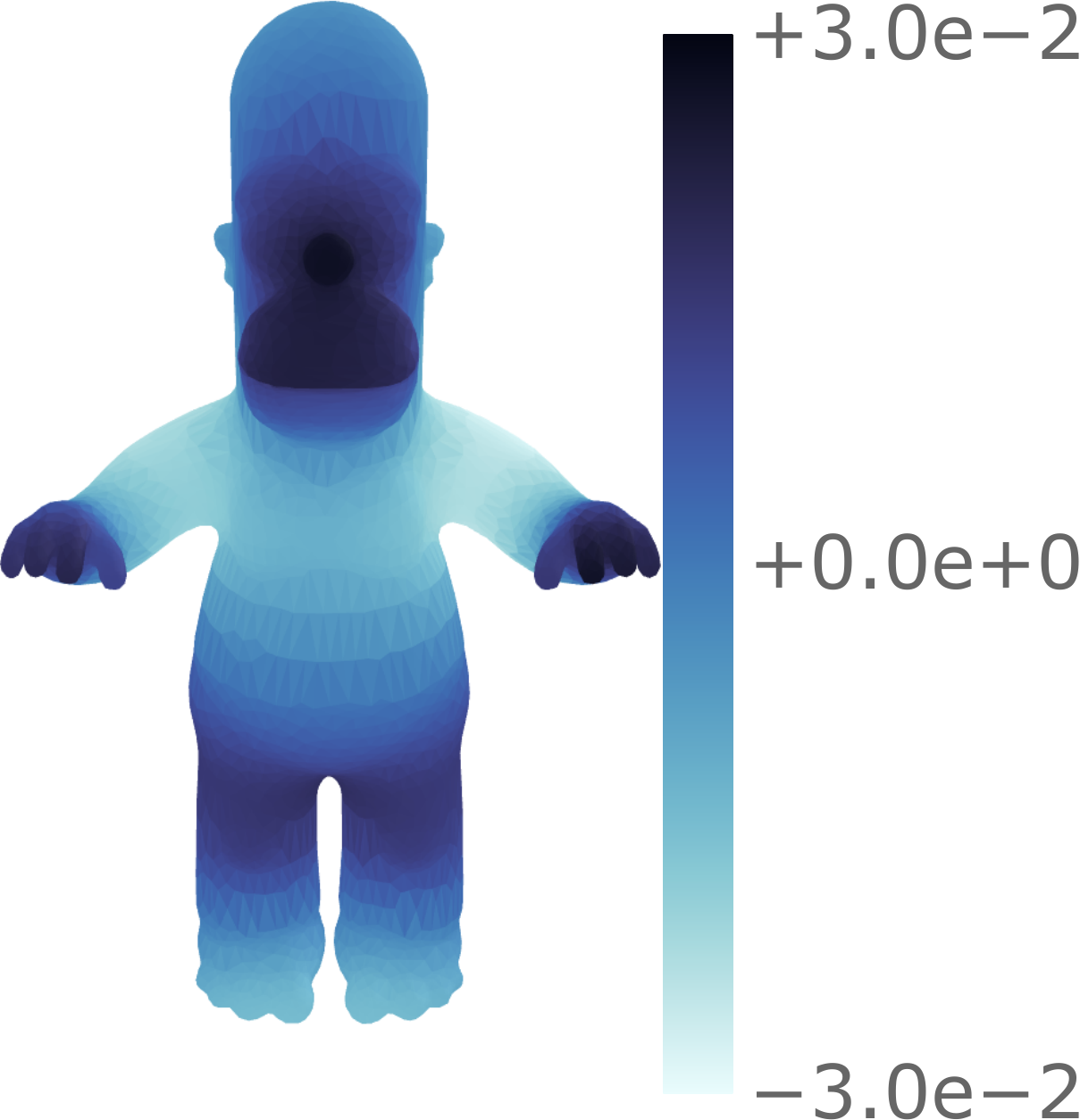}} 
	\hfill
	\subfloat[\(\mesh{\zeta_{10}}\)]
	{\includegraphics[trim={156 8 21 6},clip,width=.25\columnwidth]{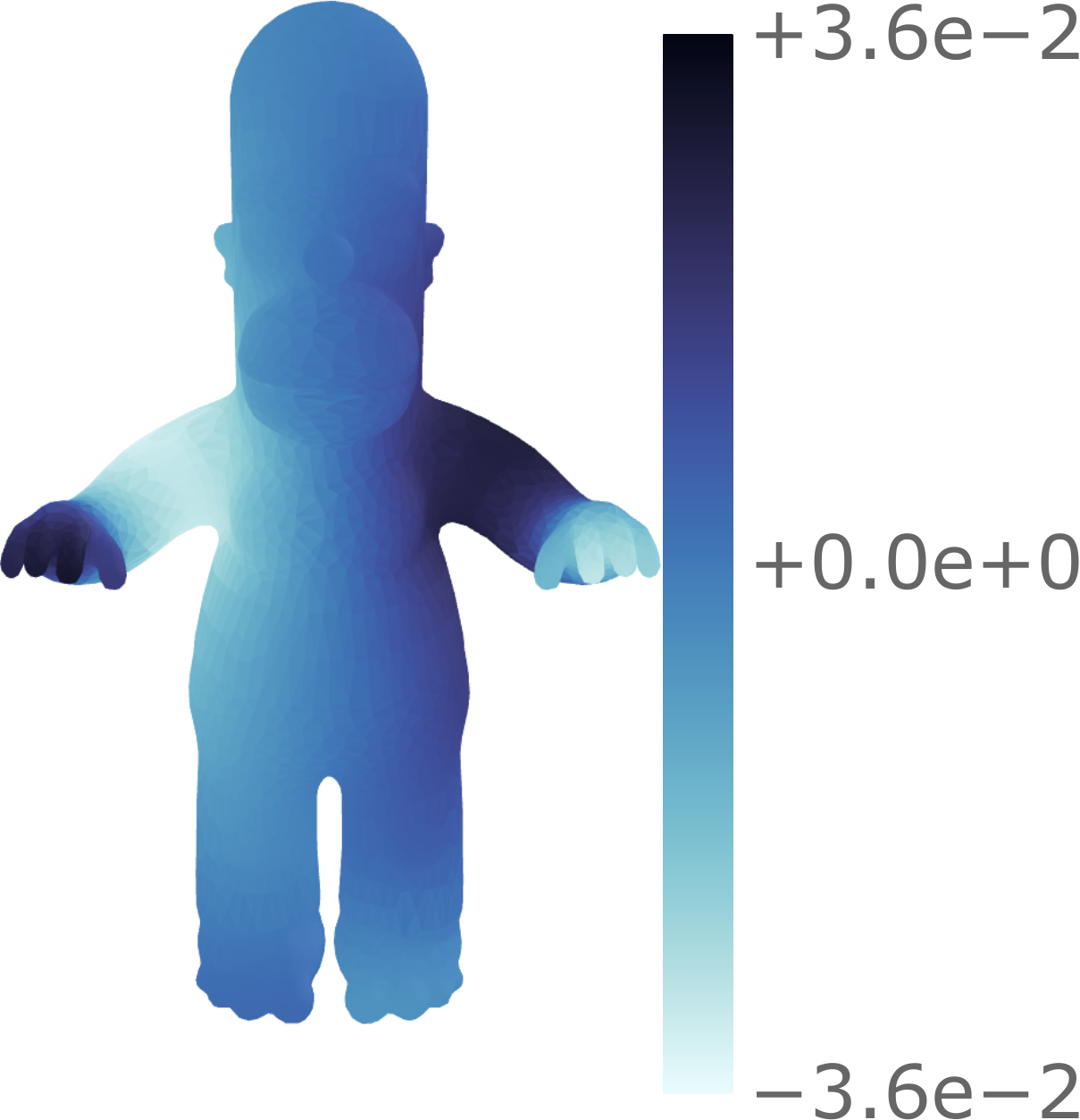}} 
	\caption{
		The third to tenth eigenvectors of the mesh Laplacian of a Homer Simpson mesh ordered by increasing eigenvalue (frequency).
		In total \(\imax=\num{1275}\) basis functions of the \(\num{5103}\) vertex mesh were computed.
		Whilst the eigenvectors are defined on the vertices, the values have been averaged onto the faces for the plot.
	}\label{fig:eigenhomers}
\end{figure}

A discrete metric
\begin{equation}
	\ell_{i j}
	= \norm{\vb*{x}_{i} - \vb*{x}_{j}}
\end{equation}
is induced by an embedding of the mesh, which changes \cref{eq:mesh_laplace_weights} to
\begin{equation}\label{eq:cotangent_laplace}
	w_{i j}
	= (\cot{\alpha_{i j}} + \cot{\beta_{i j}}) \Big/ 2,
\end{equation}
where \(\alpha_{i j}\) and \(\beta_{i j}\) are the internal angles opposite to the corresponding edge \(\ell_{i j}\), as shown in \cref{fig:mesh_laplace}.
To prove this, recall the law of cosines
\begin{equation}\label{eq:cosine}
	\cos{\alpha_{i j}}
	= -\ell_{i j}^{2} + \ell_{j k}^{2} + \ell_{i k}^{2} \Big/ 2\ell_{j k}\ell_{i k},
\end{equation}
and the usual expression for the area of a triangle
\begin{equation}\label{eq:sine}
	\sin{\alpha_{i j}}
	= 2a_{i j k} \Big/ \ell_{j k} \ell_{i k}.
\end{equation}
\cref{eq:cosine,eq:sine} may then be combined to find an expression for the cotangent of the interior angle
\begin{equation}
	\cot{\alpha_{i j}}
	= \cos{\alpha_{i j}} \Big/ \sin{\alpha_{i j}}
	= -\ell_{i j}^{2} + \ell_{j k}^{2} + \ell_{i k}^{2} \Big/ 4a_{i j k}.
\end{equation}
Through an analogous expression for the interior angle \(\beta_{i j}\), \cref{eq:cotangent_laplace} follows from \cref{eq:mesh_laplace_weights}, as elaborated in~\cite{Meyer2003}.
This cotangent Laplacian is used extensively in computer graphics~\cite{Pinkall1993}, and in the rest of this article.
A set of mesh Laplacian eigenvectors for a given mesh is presented in \cref{fig:eigenhomers}.

\subsection{Problem Formulation}\label{sec:problem_formulation}

This work generalises Slepian scale-discretised wavelets on the sphere~\cite{Roddy2022} to arbitrary manifolds.
Here, the basis functions of the Slepian concentration problem are built from the eigenfunctions of the Laplace-Beltrami operator/eigenvectors of the graph Laplacian, as opposed to the spherical harmonics.
The wavelets are constructed through a tiling of the Slepian line (akin to the harmonic line) analogously to~\cite{Wiaux2008,McEwen2018}.
The sifting convolution~\cite{Roddy2021} allows one to perform convolutions where both inputs are directional, whilst the output remains on the sphere.
By leveraging this convolution, the wavelet transforms in the Slepian domain are built on a region of the sphere (\cf{} manifold) to ensure exact reconstruction.
Convolutions are typically not well-defined on manifolds.
The sifting convolution, however, is a product in Fourier space and, as such, can be applied effectively to manifolds.
Note that performing convolutions on manifolds has been considered before; for example, in geometric deep learning, where convolutional neural network (CNN) models have been generalised to graphs by leveraging convolutions in the spectral-domain~\cite{Bruna2014,Henaff2015,Defferrard2016}, which are analogous to the sifting convolution.

In \cref{sec:working_with_manifolds}, the Slepian spatial-spectral concentration problem is presented in the manifold setting --- in particular, the spatial concentration of a bandlimited function.
The sifting convolution is then defined similarly, allowing for wavelet transforms on the manifold to be constructed.
Whilst all formulae are given explicitly in the manifold setting (through continuous integrals), they can be trivially extended to graphs through summations.
Moreover, the eigenvectors of the graph Laplacian are real, and hence the complex conjugate operators are superfluous.

\section{Working with Manifolds}\label{sec:working_with_manifolds}

In this section, a review of the Slepian spatial-spectral concentration problem adapted to manifolds is presented in \cref{sec:slepian_concentration_problem_manifolds}.
A review of the sifting convolution follows in \cref{sec:sifting_convolution_manifolds} which allows one to perform convolutions on manifolds.

\subsection{Slepian Concentration Problem on Manifolds}\label{sec:slepian_concentration_problem_manifolds}

\begin{figure}
	\centering
	\includegraphics[width=.62\columnwidth]{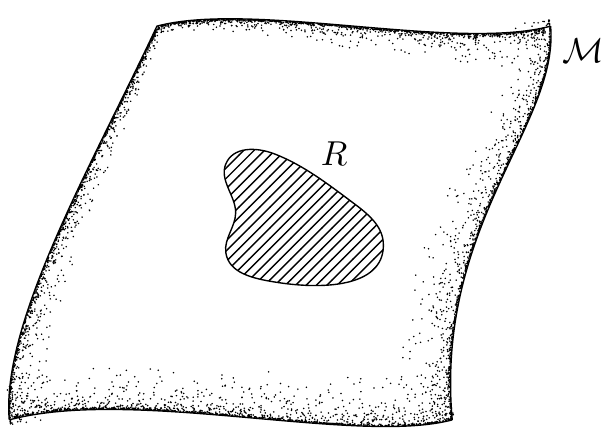}
	\caption{
		In various domains, data are observed on a partial region of a manifold \(\mathcal{M}\), such as \(R\).
	}\label{fig:region}
\end{figure}

The problem of spatial concentration of bandlimited signals (likewise spectral concentration of spacelimited signals) was first studied by Slepian, Landau and Pollak in the 1960s~\cite{Slepian1961,Landau1961,Landau1962}.
Initially considered for time-domain signals, it has since been extended to other domains such as signals on the sphere~\cite{Simons2006,Roddy2022,Xu1983,Wieczorek2005}.
This work considers optimally concentrated functions within a region \(R\) of a manifold \(\manifold{}\), an example of which is shown in \cref{fig:region}.

\subsubsection{Spatial Concentration of a Bandlimited Function}

To maximise the spatial energy concentration of a bandlimited signal \(f \in \hilbert{\manifold}\), the following ratio must be maximised
\begin{equation}
	\mu
	= \frac{\integrateRegion{\meshVolume} \abs{\mesh{f}}^{2}}{\integrateManifold{x} \abs{\mesh{f}}^{2}},
\end{equation}
where spatial concentration is measured by \(0 < \mu < 1\).
In Fourier space this can be simplified to a sum
\begin{equation}\label{eq:spatial_concentration_ratio}
	\mu
	= \frac{\meshSum f_{i} \meshYSum D_{i,j} \conj{f_{j}}}{\meshSum \abs{f_{i}}^{2}},
\end{equation}
where
\begin{equation}
	D_{i,j}
	= \integrateRegion{\meshVolume \mesh{\zeta_{i}} \mesh{\conj{\zeta_{j}}}}
\end{equation}
is an \(\imax \times \imax{}\) matrix including all basis functions of the manifold.
\cref{eq:spatial_concentration_ratio} can be rewritten as a matrix variational problem
\begin{equation}
	\mu
	= \frac{\hermitian{\vb*{f}} \vb*{D} \vb*{f}}{\hermitian{\vb*{f}} \vb*{f}},
\end{equation}
where \(\vb*{f}\) are the Fourier coefficients of \(\mesh{f}\), and are the solutions to the eigenproblem
\begin{equation}\label{eq:eigenproblem}
	\vb*{D}\vb*{f}
	= \mu \vb*{f}.
\end{equation}
The eigenvalues \(\slepian{\mu}\) act as a measure of the relative spatial concentration \(1 > \mu_{1} \geq \mu_{2} \geq \ldots \geq \mu_{\imax} > 0\), of the corresponding eigenvectors \(\vb*{f}_{1},\ \vb*{f}_{2},\ \ldots,\ \vb*{f}_{\imax}\). 
No bandlimited function can be restricted exactly within a region \(R\) and hence \(\mu_{1}<1\).
Due to the positive definiteness of \(\vb*{D}\), the smallest eigenvalue \(\mu_{\imax}\) is strictly greater than zero.
A manifold analogue of the Shannon number can be constructed by
\begin{equation}\label{eq:shannon}
	N
	= \frac{A_{R}}{A_{\manifold}} \imax,
\end{equation}
where \(A\) denotes the area of the region \(R\)/manifold \(\manifold{}\).
The Shannon number is a good estimate of the number of significant eigenvalues~\cite{Percival1993}.

\subsubsection{Slepian Decomposition}

The Slepian functions provide an alternative orthogonal basis of the manifold, decomposing a function \(f \in \hilbert{\manifold}\) into this basis
\begin{equation}
	\mesh{f}
	= \sum\limits_{p=1}^{\imax} \slepian{f} \mesh{\slepian{S}},
\end{equation}
where \(\mesh{\slepian{S}}\) are the Slepian functions and the sum is over all basis functions of the manifold.
For a well-localised function in the region \(R\) (\ie{} \(f \in \hilbert{R}\)) the sum may be truncated at the Shannon number
\begin{equation}
	\mesh{f}
	\approx \sum\limits_{p=1}^{N} \slepian{f} \mesh{\slepian{S}}
	= \slepianSum \slepian{f} \mesh{\slepian{S}},
\end{equation}
where the last line introduces a shorthand notation.
The Slepian coefficients \(\slepian{f}\) are calculated through the usual projection on to the basis functions
\begin{equation}
	\slepian{f}
	= \integrateManifold{x} \mesh{f} \mesh{\conj{\slepian{S}}}.
\end{equation}

The Slepian coefficients of a well-localised function may be computed with an integral over the region \(R\) rather than an integral over the whole manifold \(\manifold{}\)
\begin{equation}
	\slepian{f}
	\approx \frac{1}{\slepian{\mu}} \integrateRegion{\meshVolume} \mesh{f} \mesh{\conj{\slepian{S}}},
\end{equation}
as
\begin{align}
    \integrateRegion{\meshVolume} \mesh{f} \mesh{\conj{\slepian{S}}}
    & \approx \slepianSum['] \slepian[']{f} \integrateRegion{\meshVolume} \mesh{\slepian[']{S}} \mesh{\conj{\slepian{S}}} \nonumber{} \\
    &= \slepianSum['] \slepian[']{f} \hermitian{\slepian{\vb*{S}}} \vb*{D} \slepian[']{\vb*{S}}
    = \slepian{f} \slepian{\mu},
\end{align}
where \(\slepian{\vb*{S}}\) are the Fourier coefficients of \(\mesh{\slepian{S}}\).
However, there will always be a small amount of signal leakage out of the region.
Note the use of the orthogonality results
\begin{equation}\label{eq:orthogonality_sphere}
	\integrateManifold{x} \mesh{\slepian{S}} \mesh{\conj{\slepian[']{S}}}
	= \hermitian{\slepian[']{\vb*{S}}} \slepian{\vb*{S}}
	= \delta_{pp'},
\end{equation}
and
\begin{equation}
	\integrateRegion{\meshVolume} \mesh{\slepian{S}} \mesh{\conj{\slepian[']{S}}}
	= \hermitian{\slepian[']{\vb*{S}}} \vb*{D} \slepian{\vb*{S}}
	= \slepian{\mu} \hermitian{\slepian[']{\vb*{S}}} \slepian{\vb*{S}}
	= \slepian{\mu} \delta_{pp'}.
\end{equation}

One may transform from Slepian coefficients to the Fourier coefficients of the manifold by
\begin{equation}\label{eq:slepian_to_harmonic}
	f_{i}
	= \integrateManifold{x} \mesh{f} \mesh{\conj{\zeta_{i}}}
	= \slepianSum \slepian{f} {(\slepian{S})}_{i},
\end{equation}
where \({(\slepian{S})}_{i}\) are the eigenvectors of the eigenproblem \cref{eq:eigenproblem}
\begin{equation}
	{(\slepian{S})}_{i}
	= \integrateManifold{x} \mesh{\slepian{S}} \mesh{\conj{\zeta_{i}}}.
\end{equation}
The inverse operation of \cref{eq:slepian_to_harmonic} is
\begin{equation}
	\slepian{f}
	= \integrateManifold{x} \mesh{f} \mesh{\conj{\slepian{S}}}
	= \meshSum f_{i} \conj{(\slepian{S})}_{i}.
\end{equation}

\subsection{Sifting Convolution on Manifolds}\label{sec:sifting_convolution_manifolds}

A central part of wavelet transforms is the convolution.
On \(\mathbb{R}^{d}\), the convolution of a signal \(f \in \hilbert{\mathbb{R}^{d}}\) with a filter \(g \in \hilbert{\mathbb{R}^{d}}\) is defined by translating \(g\) against \(f\); however, general manifolds do not have well-defined translations.
Introduced by the authors of the current article, the sifting convolution~\cite{Roddy2021} is built on a translation that simply involves a product of the basis functions in Fourier space.
Initially defined in the spherical setting, the sifting convolution can be arbitrarily extended to other domains, including manifolds.

The complex exponentials \(\xi_{u}(x) = \exp(iux)\), with \(x,\ u \in \mathbb{R}\) form the standard orthonormal basis in the Euclidean setting.
A translation is defined through a shift of coordinates where \(\xi_{u}(x + x') = \xi_{u}(x') \xi_{u}(x)\), with \(x' \in \mathbb{R}\) and where the standard rule for exponents leads to the final equality.
A translation operator on the manifold may be defined analogously by
\begin{equation}
	\mesh{(\translation{y}\zeta_{i})}
	\equiv \meshY{\zeta_{i}} \mesh{\zeta_{i}},
\end{equation}
where \(y\) is a point on the manifold.
A natural way to define the translation of an arbitrary function \(f \in \hilbert{\manifold}\) is thus
\begin{equation}\label{eq:translation_mesh}
	\mesh{(\translation{y}f)}
	= \meshSum f_{i} \meshY{\zeta_{i}} \mesh{\zeta_{i}},
\end{equation}
which implies
\begin{equation}
	{(\translation{y}f)}_{i}
	= f_{i} \meshY{\zeta_{i}}.
\end{equation}

The sifting convolution on the manifold of \(f,\ g \in \hilbert{\manifold}\) follows in the usual manner by the inner product
\begin{equation}
	\mesh{\convolution{f}{g}}
	\equiv \braket*{\translation{x}f}{g},
\end{equation}
which is a product in Fourier space
\begin{equation}\label{eq:harmonic_convolution}
	{\convolution{f}{g}}_{i}
	= f_{i} \conj{g_{i}},
\end{equation}
as
\begin{align}
	\mesh{\convolution{f}{g}}
	 & = \braket*{\translation{x}f}{g} \nonumber{}                                                                                           \\
	 & = \integrateManifold{y} \meshY{(\translation{x}f)} \meshY{\conj{g}} \nonumber{}                                                       \\
	 & = \integrateManifold{y} \meshSum f_{i} \mesh{\zeta_{i}} \meshY{\zeta_{i}} \meshYSum \conj{g_{j}} \meshY{\conj{\zeta_{j}}} \nonumber{} \\
	 & = \meshSum f_{i} \conj{g_{i}} \mesh{\zeta_{i}}.
\end{align}
Note that the Fourier representation of the convolution is a product and, hence, is efficient to compute.
It can be shown that the translation operator is a sifting convolution of a function with the shifted Dirac delta function
\begin{align}
	\mesh{\convolution{f}{\delta_{y}}}
	 & = \meshSum f_{i} \conj{{(\delta_{y})}_{i}} \mesh{\zeta_{i}} \nonumber{} \\
	 & = \meshSum f_{i} \meshY{\zeta_{i}} \mesh{\zeta_{i}} \nonumber{}         \\
	 & = \mesh{(\translation{y}f)},
\end{align}
where the first and last lines follow by \cref{eq:harmonic_convolution,eq:translation_mesh} respectively.
The sifting convolution and translation of the manifold are therefore natural analogues of their respective Euclidean operators.
In \cref{sec:slepian_wavelets}, this convolution is leveraged to define wavelets restricted to a region of the manifold using the basis functions introduced in \cref{sec:slepian_concentration_problem_manifolds}.

\section{Slepian Wavelets}\label{sec:slepian_wavelets}

Here, the Slepian wavelet theory on manifolds is developed.
Initially, \cref{sec:slepian_sifting_convolution} extends the sifting convolution~\cite{Roddy2021} to the Slepian domain.
With a convolution to hand, the Slepian wavelet transform is developed in \cref{sec:slepian_scale_discretised_wavelets} --- where an appropriate admissibility condition must hold for exact reconstruction.
The generating functions that define the wavelets are presented in \cref{sec:generating_functions}.
Lastly, some properties of the wavelets are discussed in \cref{sec:properties}.
The construction here is the same as a previous work by the authors of the current article~\cite{Roddy2022}, where now the integrals are over the manifold (graph) rather than strictly the sphere.
For brevity, the formulae will be presented just in real space and not also in Fourier space.

\subsection{Slepian Sifting Convolution}\label{sec:slepian_sifting_convolution}

To construct wavelets in a region of the manifold, a suitable convolution is required.
The sifting convolution on the manifold, defined in \cref{sec:sifting_convolution_manifolds} and adapted from~\cite{Roddy2021}, can be extended to work with the Slepian functions as a basis.
The construction of the Slepian sifting convolution on manifolds is analogous to~\cite{Roddy2022}, but with the Slepian functions of the manifold defined in \cref{sec:slepian_concentration_problem_manifolds} (rather than those of the sphere).
Slepian wavelets may now be defined utilising this convolution in Slepian space.

\subsection{Slepian Scale-Discretised Wavelets}\label{sec:slepian_scale_discretised_wavelets}

A Slepian wavelet transform can be constructed through a tiling of Slepian space, where \(p\) is restricted to \(N=\imax A_{R}/A_{\manifold}\) (or \(\imax{}\) for the whole manifold).
Spatially localised, scale-dependent content of a signal may be probed through a scale-discretised wavelet transform.
The construction of these wavelets is analogous to~\cite{Wiaux2008,McEwen2018} but are computed in the Slepian basis rather than the basis functions of the sphere (\cf{} manifold).

For a signal of interest \(f \in \hilbert{R}\) concentrated within a region \(R\), the wavelet coefficients \(W^{\Psi^{j}} \in \hilbert{R}\) are defined through a sifting convolution of \(f\) with the wavelet \(\Psi^{j} \in \hilbert{R}\) for wavelet scale \(j\)
\begin{equation}
	\mesh{W^{\Psi^{j}}}
	= \mesh{\convolution{\Psi^{j}}{f}}
	= \integrateManifold{y} \meshY{(\translation{x}\Psi^{j})} \meshY{\conj{f}}.
\end{equation}
Wavelets are typically paired with a scaling function, each capturing different underlying scales of the signal.
Scaling coefficients \(W^{\Phi} \in \hilbert{R}\) may be similarly defined by a sifting convolution between \(f\) and the scaling function \(\Phi \in \hilbert{R}\)
\begin{equation}
	\mesh{W^{\Phi}}
	= \mesh{\convolution{\Phi}{f}}
	= \integrateManifold{y} \meshY{(\translation{x}\Phi)} \meshY{\conj{f}}.
\end{equation}
Supposing that the wavelets and scaling function satisfy an admissibility condition, a function \(f\) may be reconstructed from its wavelet and scaling coefficients by
\begin{align}\label{eq:synthesis}
	\mesh{f}
	& = \integrateManifold{y} \meshY{(\translation{x}\Phi)} \meshY{W^{\Phi\ast}} \nonumber{} \\
	& + \integrateManifold{y} \waveletSum \meshY{(\translation{x}\Psi^{j})} \meshY{W^{\Psi^{j}\ast}}.
\end{align}
The parameters \(J_{0}\) and \(J\) here represent the lowest and highest scales \(j\) of the wavelet decomposition respectively --- to ensure exact reconstruction these parameters must be set consistently.
The admissibility condition required for synthesis \cref{eq:synthesis} to hold is thus
\begin{equation}\label{eq:admissibility}
	\abs{\slepian{\Phi}}^{2}
	+ \waveletSum \abs{\slepian{\Psi}^{j}}^{2}
	= 1,\ \forall p.
\end{equation}
Wavelets and a scaling function that satisfy this admissibility condition may now be defined.

\subsection{Generating Functions}\label{sec:generating_functions}

A set of smooth generating functions is required to tile the Slepian line.
Here a set of such functions defined by~\cite{Wiaux2008} is utilised, a short summary follows.
Consider the \(C^{\infty}\) Schwartz function with compact support on \(\interval{-1}{1}\):
\begin{equation}
	s(t) \equiv
	\begin{cases}
		\exp(1/(t^{2}-1)), & t \in \interval{-1}{1}    \\
		0,                 & t \notin \interval{-1}{1}
	\end{cases}
\end{equation}
for \(t \in \mathbb{R}\).
The positive real parameter \(\lambda \in \realPosParam{}\) may then be introduced to map \(s(t)\) to
\begin{equation}
	s_{\lambda}(t)
	\equiv s\bigg(\frac{2\lambda}{\lambda-1}(t-\lambda^{-1}) - 1\bigg),
\end{equation}
which has compact support in \(\interval{1/\lambda}{1}\).
One can define the smoothly decreasing function \(k_{\lambda}\) by
\begin{equation}
	k_{\lambda}(t)
	\equiv \int\limits_{t}^{1} \dd{t'} \frac{s^{2}_{\lambda}(t')}{t'}
	\bigg/ \int\limits_{1/\lambda}^{1} \dd{t'} \frac{s^{2}_{\lambda}(t')}{t'},
\end{equation}
which is unity for \(t < 1/\lambda{}\), zero for \(t > 1\), and smoothly decreasing from unity to zero for \(t \in \interval{1/\lambda}{1}\).
The wavelet generating function is defined as
\begin{equation}
	\kappa_{\lambda}(t)
	\equiv \sqrt{k_{\lambda}(t/\lambda) - k_{\lambda}(t)},
\end{equation}
and the scaling function generating function is
\begin{equation}
	\eta_{\lambda}(t)
	\equiv \sqrt{k_{\lambda}(t)}.
\end{equation}

\begin{figure}
	\centering
	\includegraphics[width=\columnwidth]{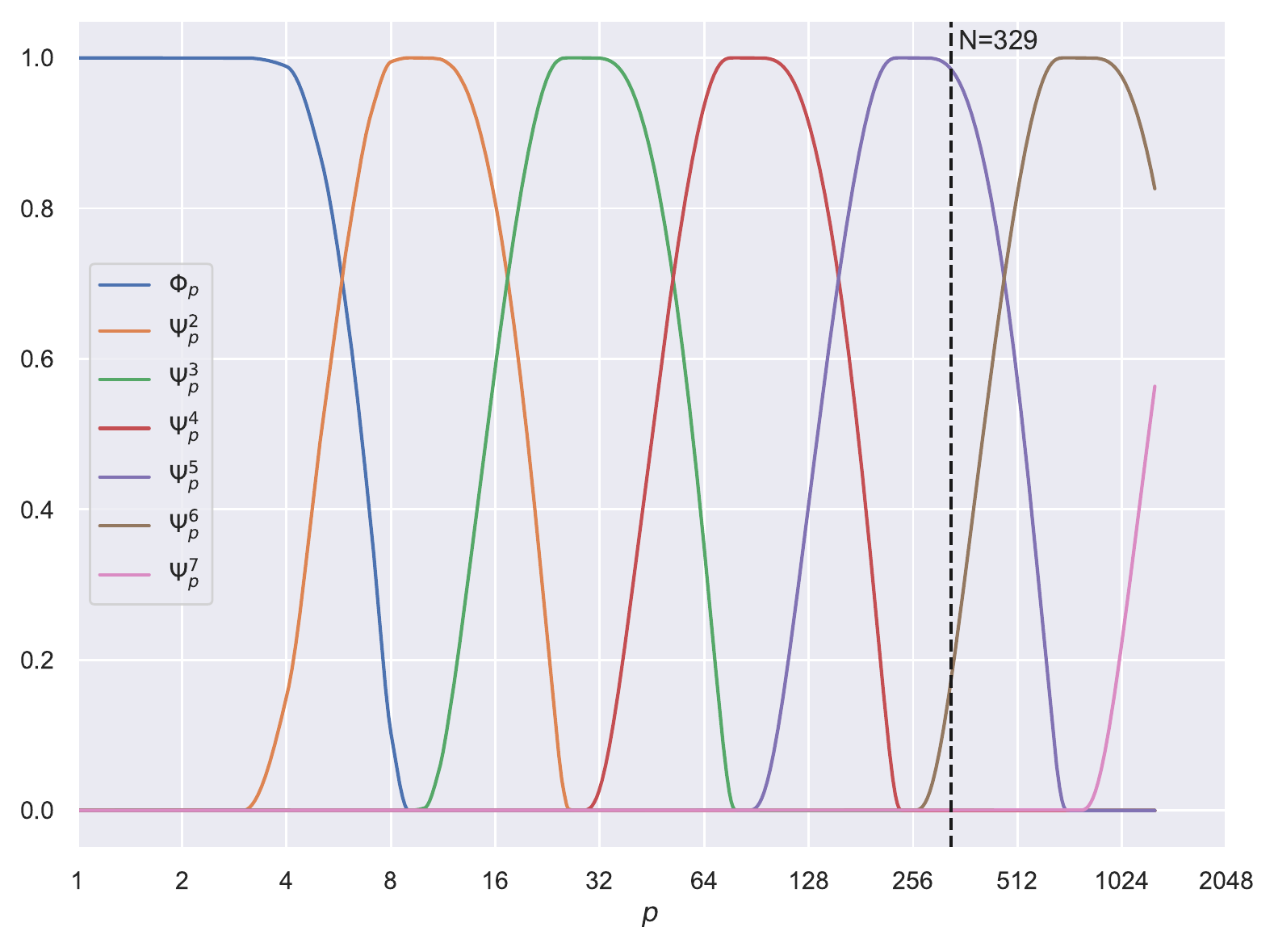}
	\caption{
		The tiling of the Slepian line with parameters \(\lambda=3\) and \(J_{0}=2\) with \(\imax=\num{1275}\) basis functions.
		The black dashed line marks the Shannon number for the Homer head region \(N=329\).
		The scaling function and the first five wavelets are non-zero, as the coefficients are within the Shannon number.
	}\label{fig:tiling}
\end{figure}

An instinctive approach is to define the wavelets \(\slepian{\Psi}^{j}\) from the generating functions \(\kappa_{\lambda}\) to have support on \(\interval{\lambda^{j-1}}{\lambda^{j+1}}\), yielding
\begin{equation}
	\slepian{\Psi}^{j}
	\equiv \kappa_{\lambda}\bigg(\frac{p}{\lambda^{j}}\bigg).
\end{equation}
For \(p \geq \lambda^{J_{0}}\) the admissibility condition \cref{eq:admissibility} is satisfied for these wavelets, where \(J_{0}\) is the lowest wavelet scale used in the decomposition.
Modes that cannot be probed by wavelets can be extracted through the construction of a scaling function \(\Phi{}\) (\ie{} modes with \(p < \lambda^{J_{0}}\))
\begin{equation}
	\slepian{\Phi}
	\equiv \eta_{\lambda}\bigg(\frac{p}{\lambda^{J_{0}}}\bigg).
\end{equation}
Exact reconstruction can be achieved by setting \(J\) appropriately
\begin{equation}
	J = \lceil{} \log_{\lambda}(N)\rceil{}.
\end{equation}
Assuming that \(0 \leq J_{0} < J\) is satisfied, the lowest wavelet scale \(J_{0}\) is arbitrary.
The Slepian wavelets are constructed by the tiling of the Slepian line as shown in \cref{fig:tiling}.

\subsection{Properties}\label{sec:properties}

The properties of Slepian wavelets on manifolds are reviewed here, which are the same as in the spherical case~\cite{Roddy2022}.
In comparison to standard scale-discretised wavelets, the properties are often similar, but not always.

\subsubsection{Localisation}\label{sec:localisation}

Standard scale-discretised wavelet constructions are built on a tiling of the harmonic line, and hence a higher value of \(j\) in the wavelets \(\Psi^{j}\) corresponds to smaller scales.
In contrast, Slepian wavelets are built on the Slepian harmonic line, and thus, \(j\) is a measure of localisation rather than scale.

\subsubsection{Wavelet Energy}

The wavelet energy is
\begin{equation}
	\norm{\Psi^{j}}^{2}
	= \slepianSum \abs{\slepian{\Psi}^{j}}^{2}.
\end{equation}
A similar expression exists for the scaling function energy.

\subsubsection{Parseval Frame}

A Parseval frame is satisfied by Slepian scale-discretised wavelets on manifolds
\begin{align}
	A\norm{f}^{2} \leq
    & \integrateManifold{x} \abs{\braket*{\translation{x}\Phi}{f}}^{2} \nonumber{} \\
	& + \waveletSum \integrateManifold{x} \abs{\braket*{\translation{x}\Psi^{j}}{f}}^{2}
	\leq B\norm{f}^{2},
\end{align}
where \(A,\ B \in \realPosParam{}\).
It can be shown that this holds for \(A = B = 1\), implying that the energy of \(f\) is conserved in wavelet space.

\subsubsection{Wavelet Domain Variance}

For notational brevity, define a quantity
\begin{equation}
	\varphi \in \set{\Phi,\Psi^{j}}
\end{equation}
to represent both the scaling function and the wavelets.
Consider homogenous and isotropic noise defined by its power spectrum
\begin{equation}
	\expval*{f_{i} \conj{f_{j}}}
	= C_{i} \delta_{i j},
\end{equation}
where \(C_{i} = \sigma^{2}\) for white noise.
The corresponding power spectrum in Slepian space is
\begin{equation}
	\expval*{\slepian{f} \conj{\slepian[']{f}}}
	= \sigma^{2} \delta_{pp'}.
\end{equation}
For the common case of zero-mean Gaussian noise, the expected value of the wavelet/scaling coefficients is zero, and hence the wavelet domain variance is
\begin{equation}
	\variance{\mesh{W^{\varphi}}}
	= \sigma^{2} \slepianSum \abs{\slepian{\varphi}}^{2} \abs{\mesh{\slepian{S}}}^{2},
\end{equation}
and, as such, the variance depends on the position on the manifold.

\section{Numerical Illustration}\label{sec:numerical_illustration}

In this section, the construction and application of Slepian wavelets for an example region on a mesh (\cf{} manifold) is demonstrated.
The Slepian functions and eigenvalues of a region of a Homer Simpson mesh are presented in \cref{sec:homer_region}.
A field is constructed on the region of the mesh in \cref{sec:wavelet_transform}, and the resulting wavelets and wavelet coefficients are computed.
A possible use of Slepian wavelets is shown in \cref{sec:wavelet_denoising} through a straightforward denoising procedure.
The wavelet generating functions, discussed in \cref{sec:generating_functions}, are constructed through the \texttt{S2LET}~\cite{Leistedt2013} code.
Further, the \texttt{SLEPLET}~\cite{Roddy2022a} code has been developed to perform the work in this article.

\subsection{Homer Region}\label{sec:homer_region}

\begin{figure}
    \hspace{20mm}
	\includegraphics[trim={156 8 21 6},clip,width=.62\columnwidth]{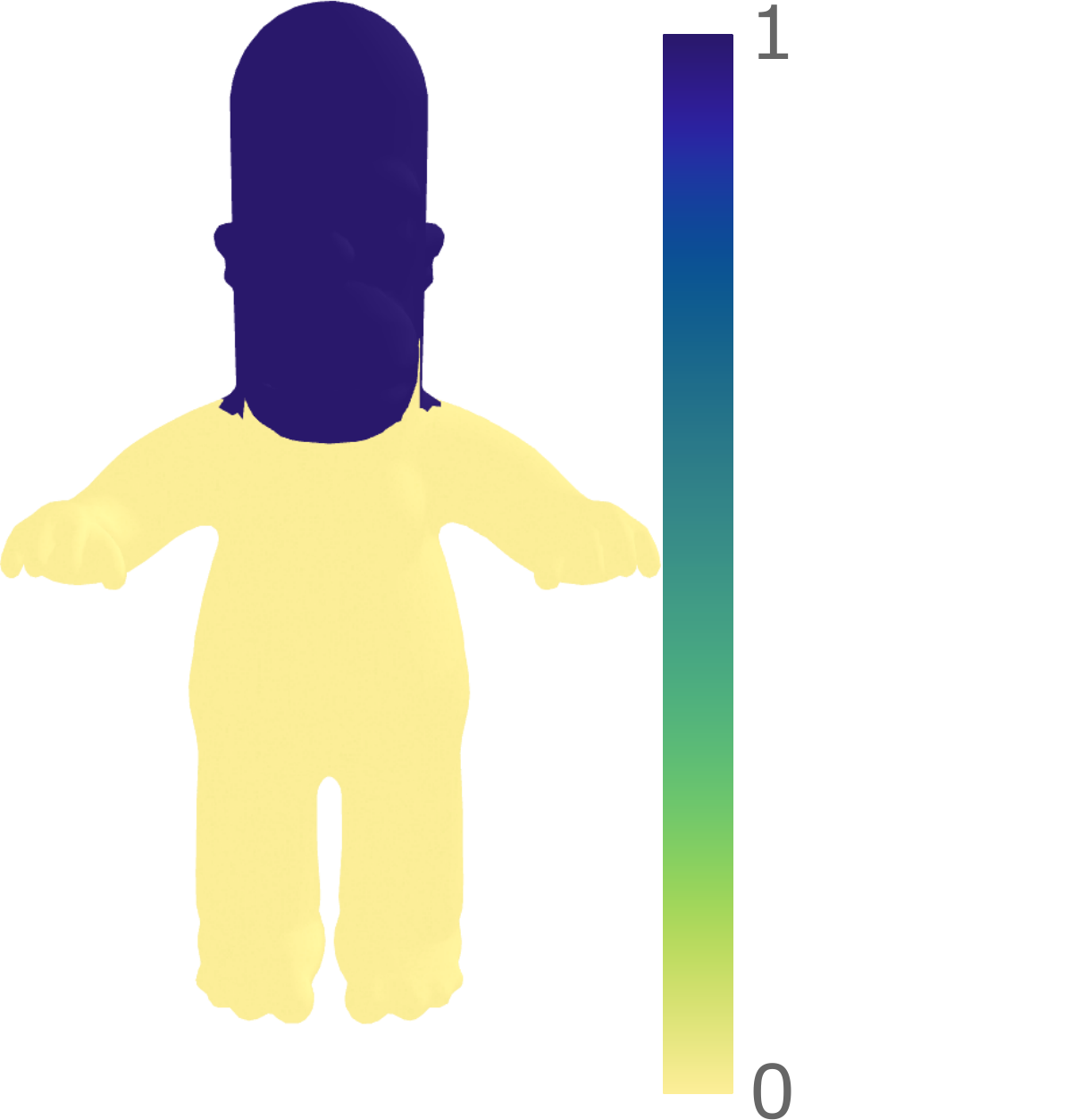}
	\caption{
		The head region (in blue) chosen to compute the Slepian functions of the Homer mesh.
	}\label{fig:homer_region}
\end{figure}

\begin{figure}
	\centering
	\subfloat[\(\mesh{S_{1}} \newline
        \mu_{1}=1.00\)]
	{\includegraphics[trim={101 0 3 3},clip,width=.33\columnwidth]{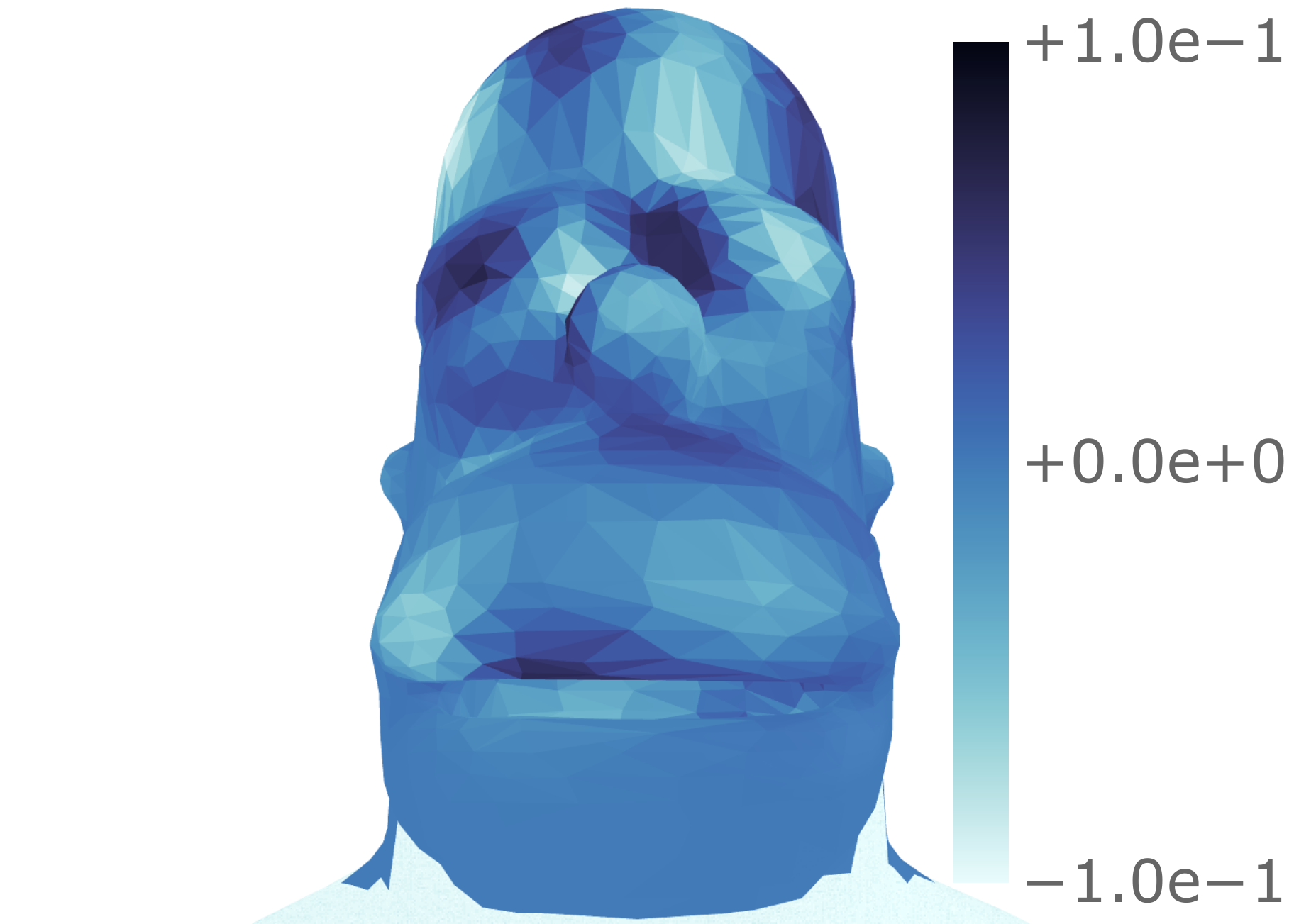}} 
	\hfill
	\subfloat[\(\mesh{S_{10}} \newline
        \mu_{10}=1.00\)]
	{\includegraphics[trim={101 0 3 3},clip,width=.33\columnwidth]{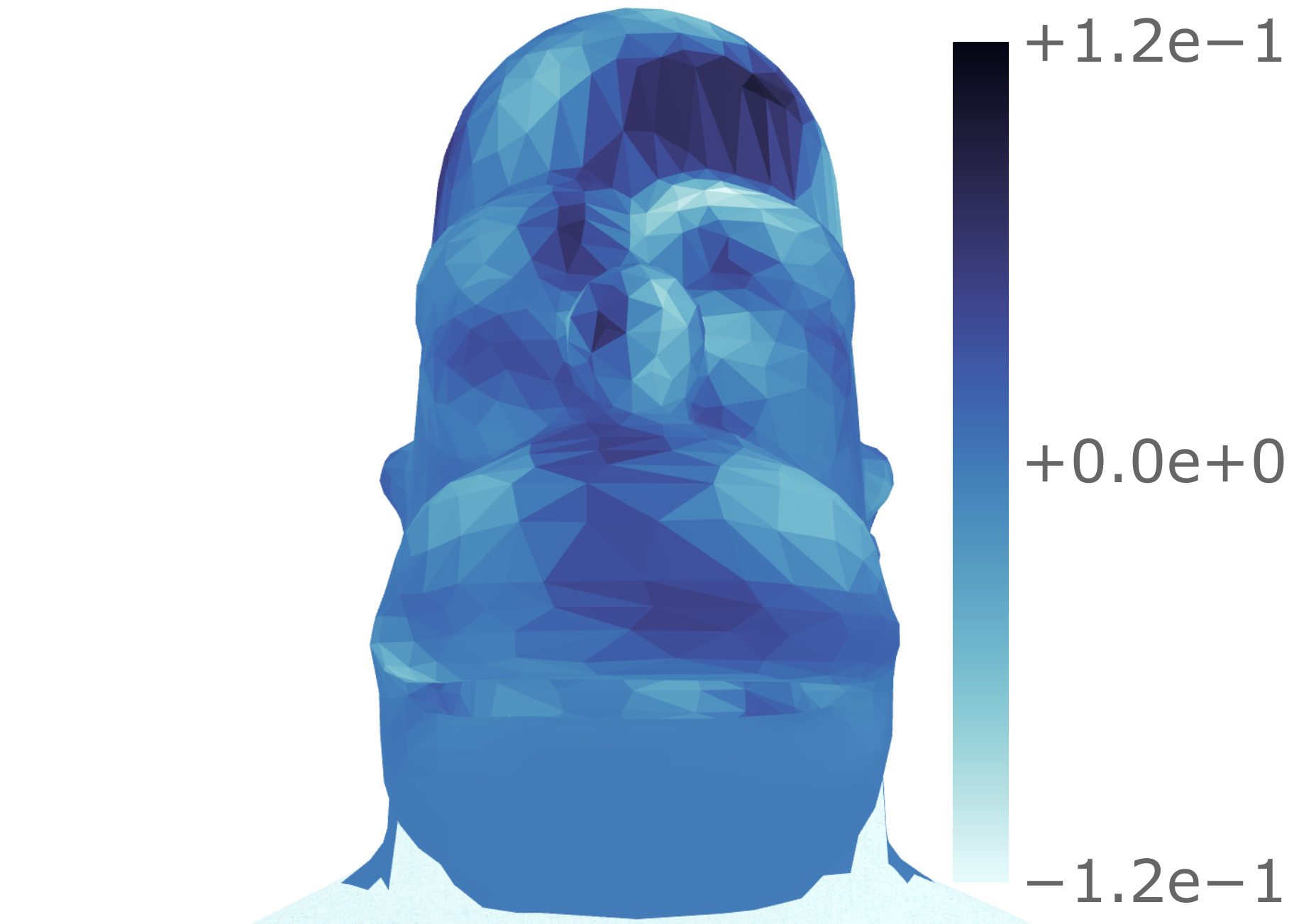}} 
	\hfill
	\subfloat[\(\mesh{S_{25}} \newline
        \mu_{25}=1.00\)]
	{\includegraphics[trim={101 0 3 3},clip,width=.33\columnwidth]{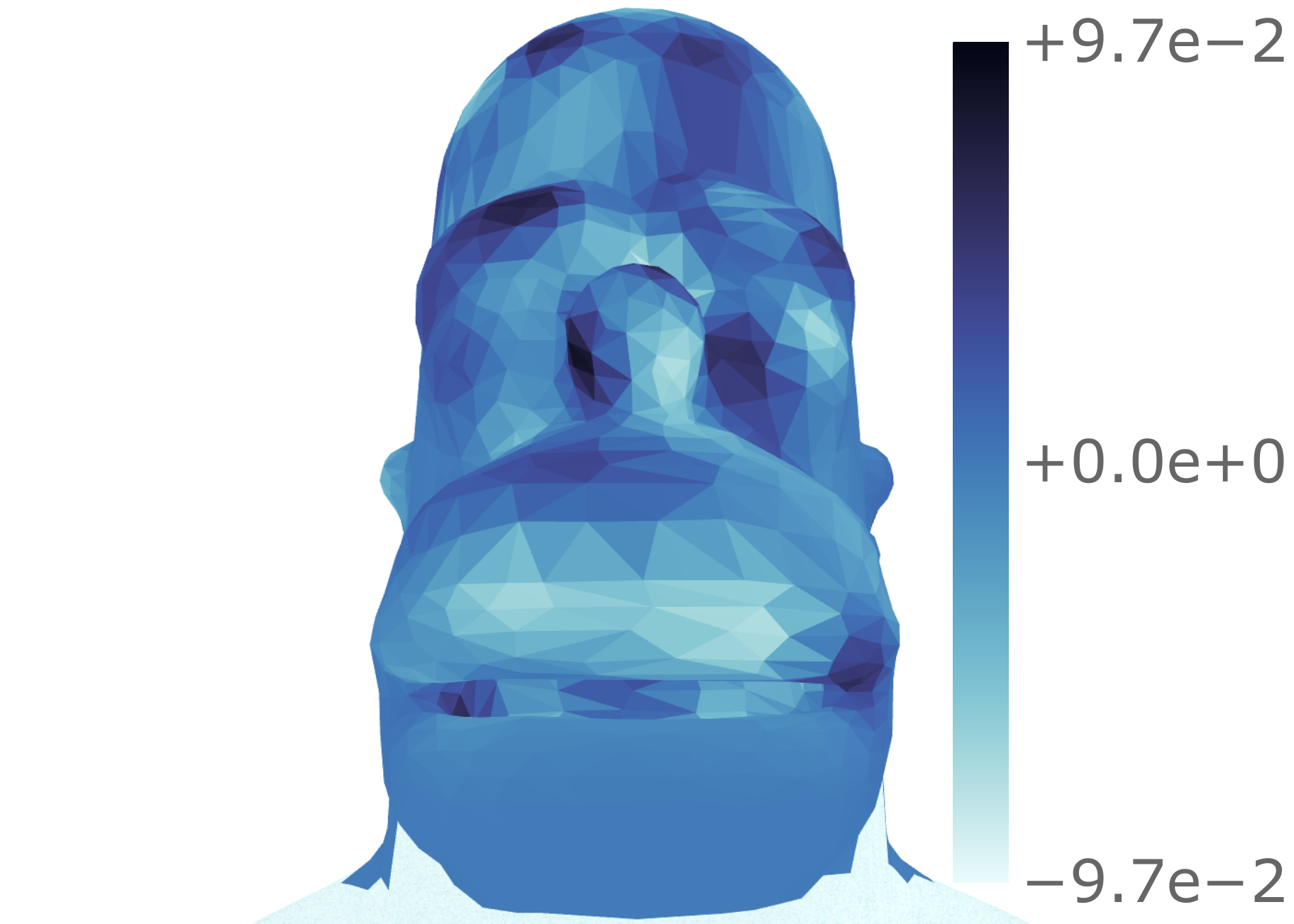}} 
	\newline
	\subfloat[\(\mesh{S_{50}} \newline
        \mu_{50}=1.00\)]
	{\includegraphics[trim={101 0 3 3},clip,width=.33\columnwidth]{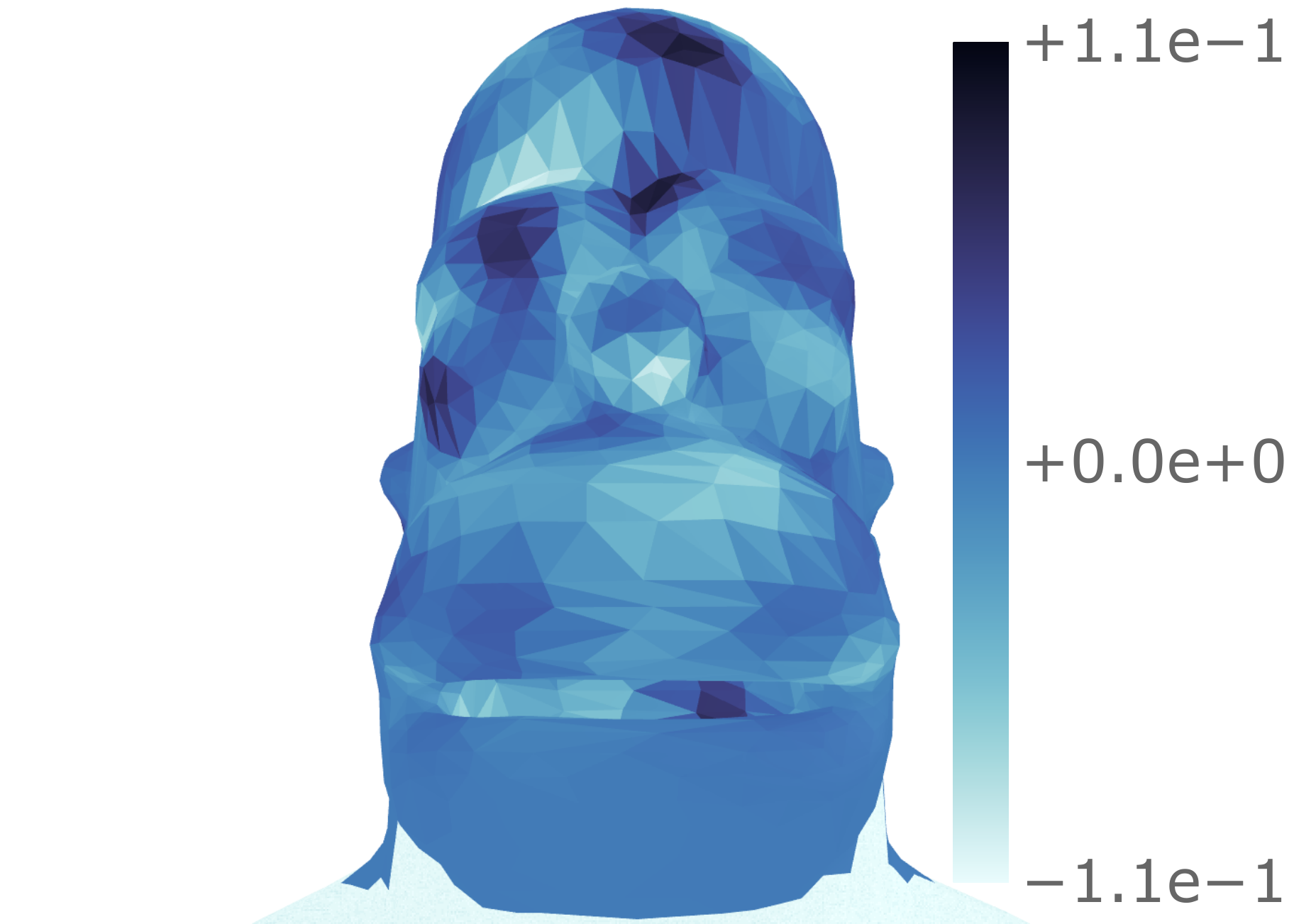}} 
	\hfill
	\subfloat[\(\mesh{S_{100}} \newline
        \mu_{100}=1.00\)]
	{\includegraphics[trim={101 0 3 3},clip,width=.33\columnwidth]{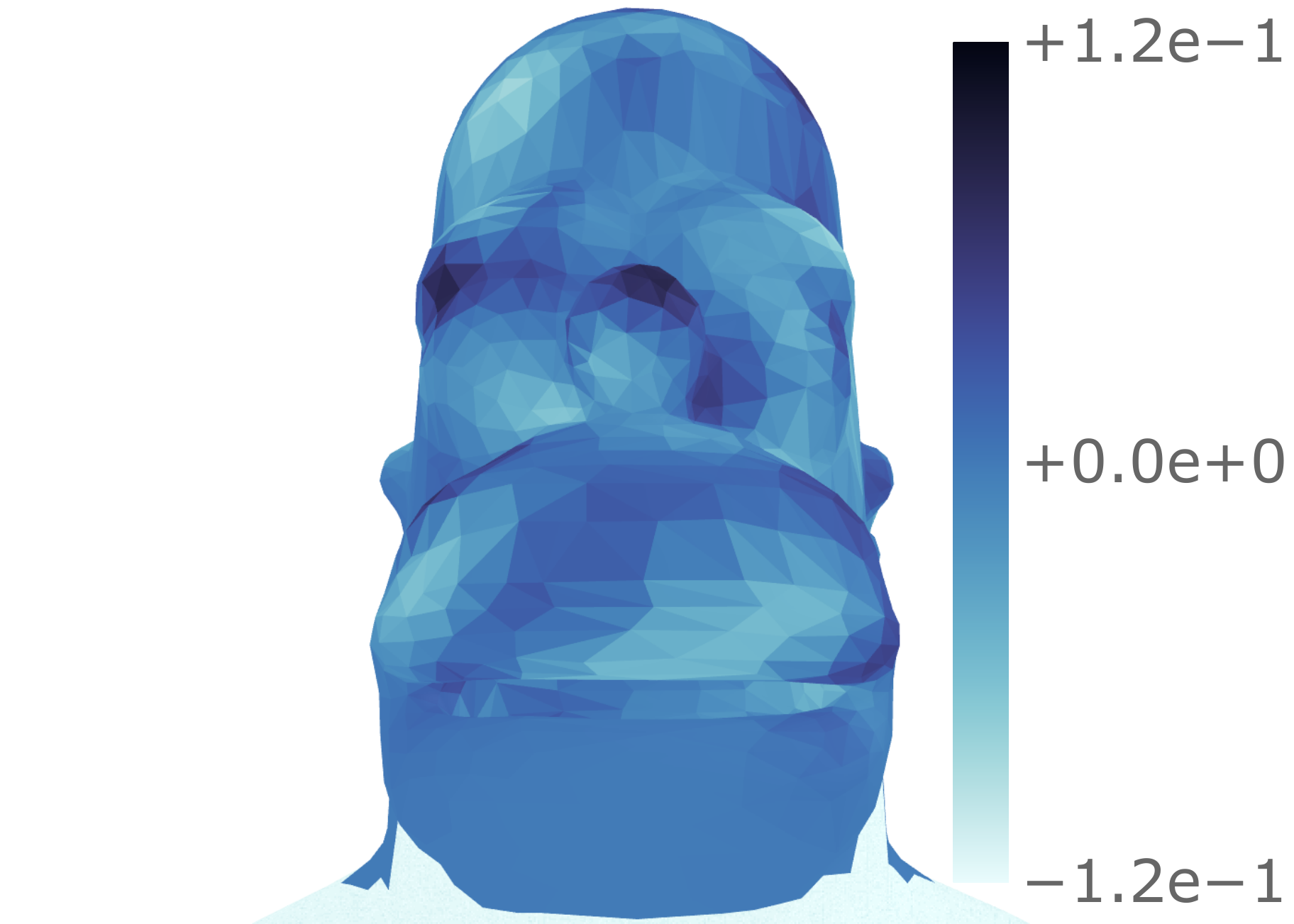}} 
	\hfill
	\subfloat[\(\mesh{S_{200}} \newline
        \mu_{200}=1.00\)]
	{\includegraphics[trim={101 0 3 3},clip,width=.33\columnwidth]{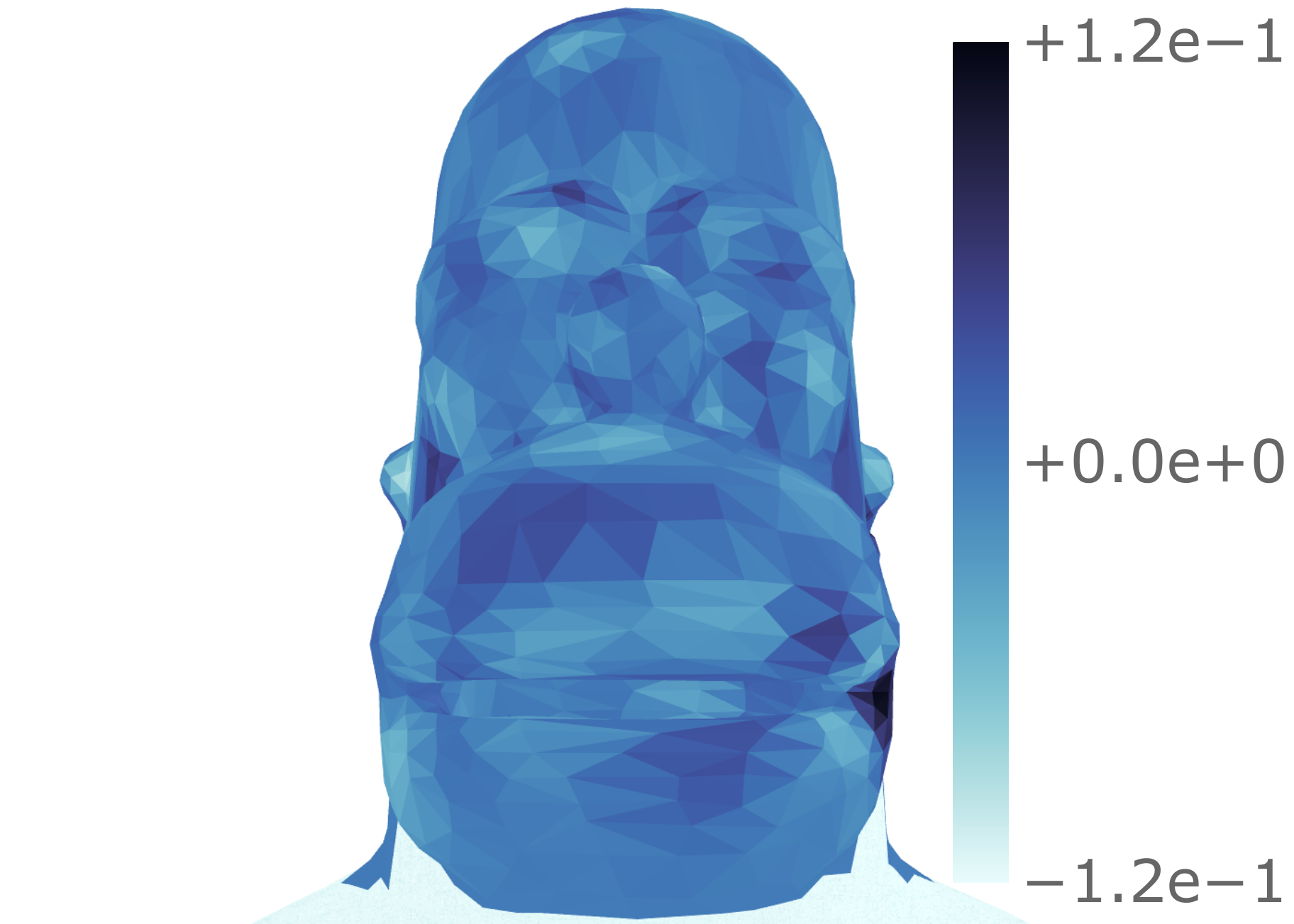}} 
	\caption{
		The Slepian functions of the Homer head region \(\mesh{\slepian{S}}\) for \(p \in \set{1, 10, 25, 50, 100, 200}\) shown left-to-right, top-to-bottom.
		The corresponding eigenvalue \(\slepian{\mu}\) is a measure of the concentration within the given region \(R\), that remains \(\almost{\num{1}}\) for many \(p\) values before decreasing towards zero.
		Whilst the Slepian functions are defined on the vertices, the values have been averaged onto the faces for the plot.
	}\label{fig:slepian_functions}
\end{figure}

\begin{figure}
	\centering
	\includegraphics[width=\columnwidth]{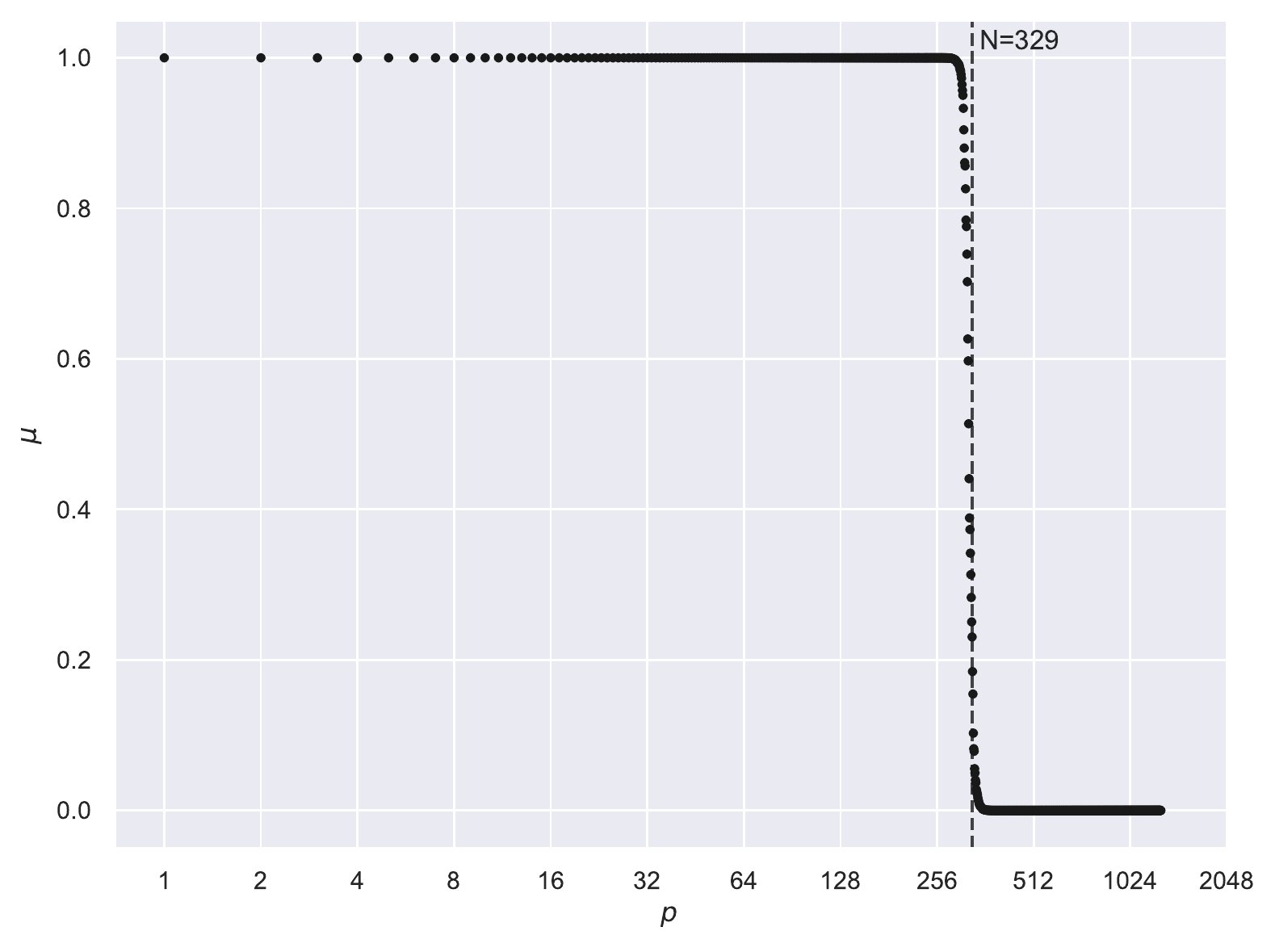}
	\caption{
		The ordered eigenvalues of the Homer head region.
		The majority of the eigenvalues are \(\almost{\num{1}}\) before decreasing rapidly towards zero around the Shannon number \(N=329\).
	}\label{fig:slepian_eigenvalues}
\end{figure}

A region of a manifold is created on a mesh of Homer Simpson; \cref{fig:homer_region} presents the masked region of Homer's head.
The Slepian functions of this region are computed by solving the eigenproblem \cref{eq:eigenproblem}, and then performing an inverse Fourier transform on the mesh.
The resulting Shannon number \cref{eq:shannon} of the region \(R\) is \(N=329\).
A set of Slepian functions of the mesh is shown in \cref{fig:slepian_functions} for \(p \in \set{1, 10, 25, 50, 100, 200}\).
Comparing panel (f) to the earlier panels, one can see that the Slepian functions become more spread out in the region, representing worse concentration.
The corresponding eigenvalues \(\slepian{\mu}\) are a measure of spatial concentration, which remain \(\almost{\num{1}}\) for many \(p\) values before rapidly decreasing towards zero around \(N\).
The eigenvalues of this Homer region are shown in \cref{fig:slepian_eigenvalues}.

\subsection{Wavelet Transform}\label{sec:wavelet_transform}

\begin{figure}
	\centering
	\subfloat[\(\mesh{\Phi}\)]
	{\includegraphics[trim={101 0 3 3},clip,width=.33\columnwidth]{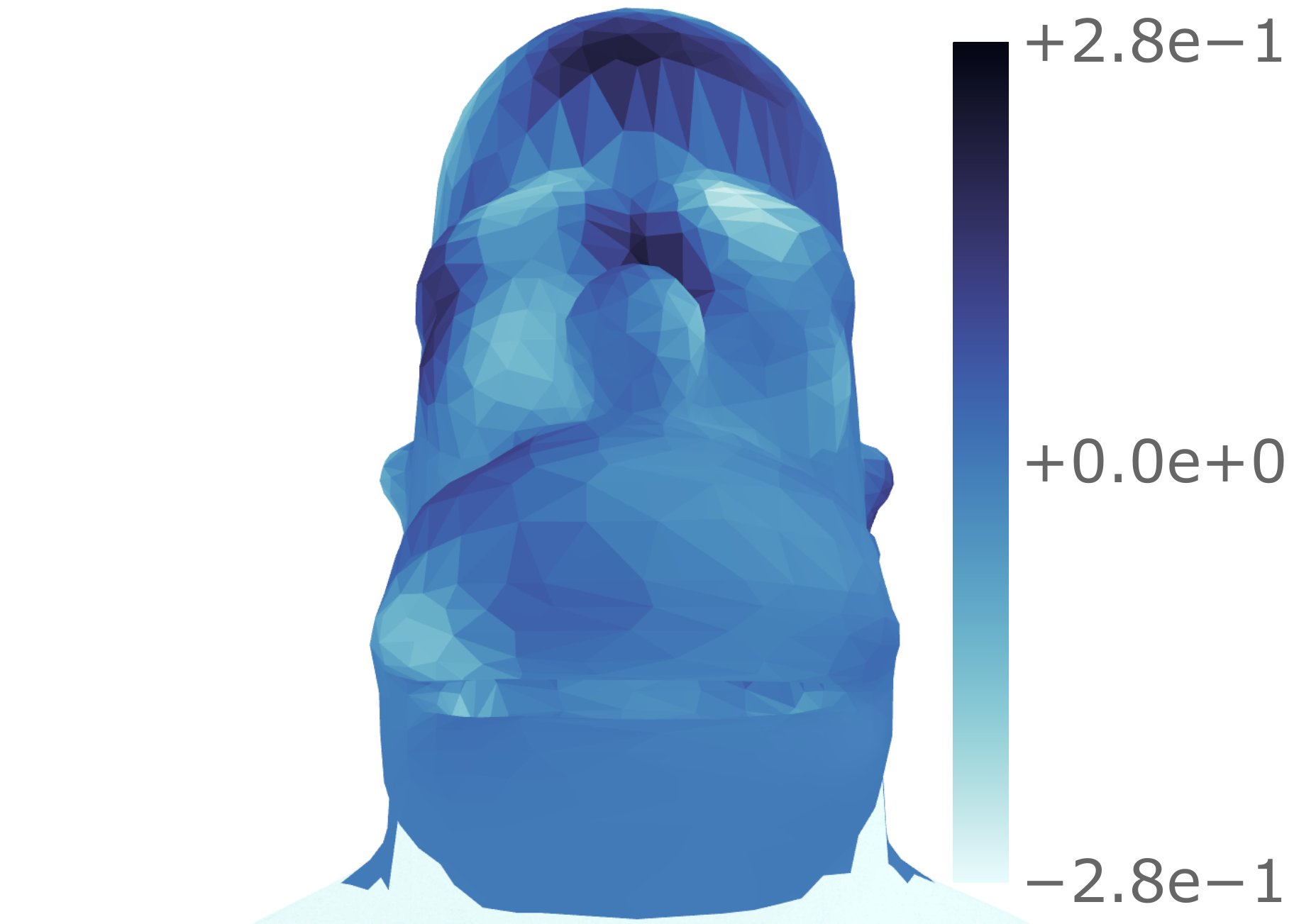}}
	\hfill
	\subfloat[\(\mesh{\Psi^{2j}}\)]
	{\includegraphics[trim={101 0 3 3},clip,width=.33\columnwidth]{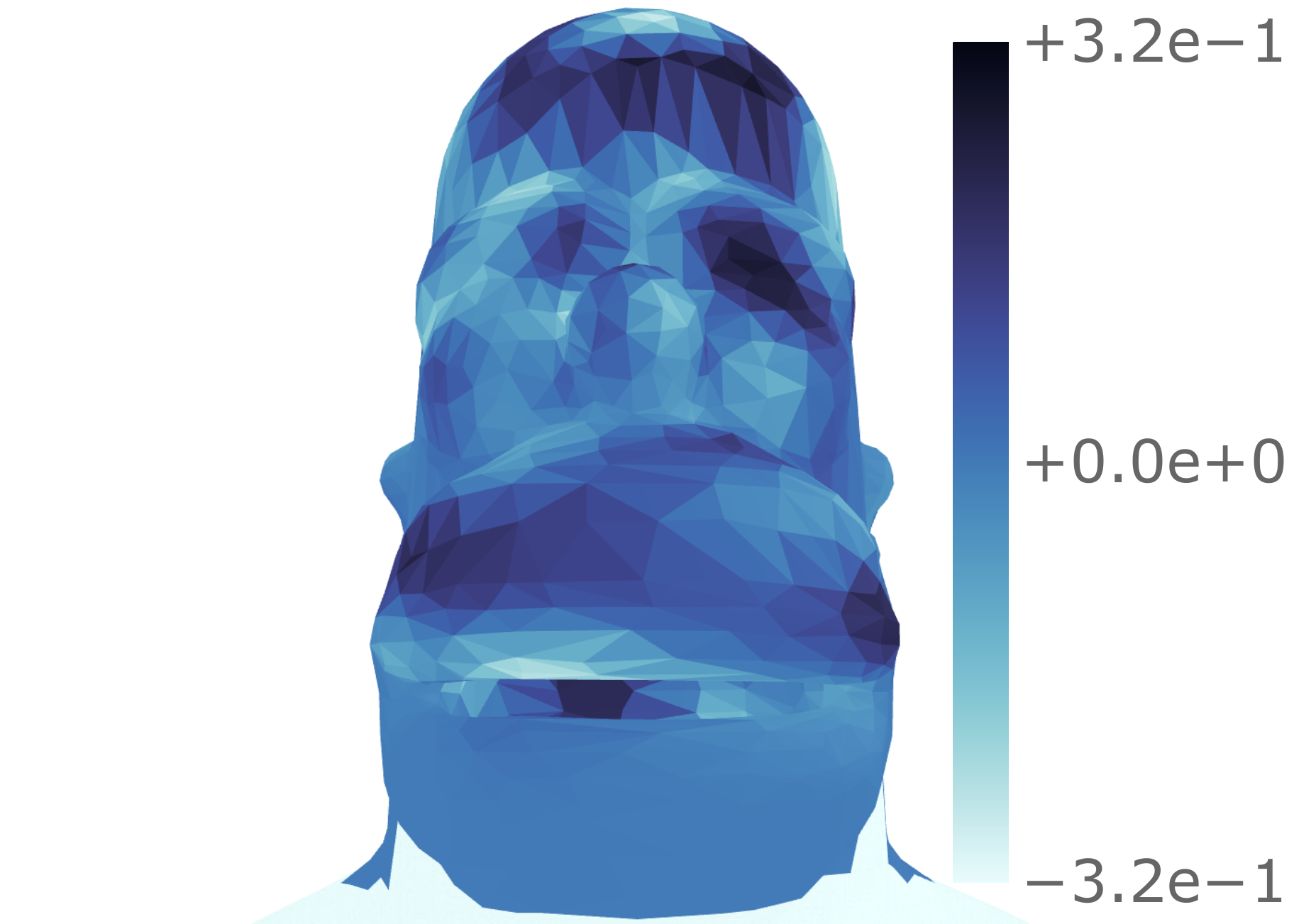}}
	\hfill
	\subfloat[\(\mesh{\Psi^{3j}}\)]
	{\includegraphics[trim={101 0 3 3},clip,width=.33\columnwidth]{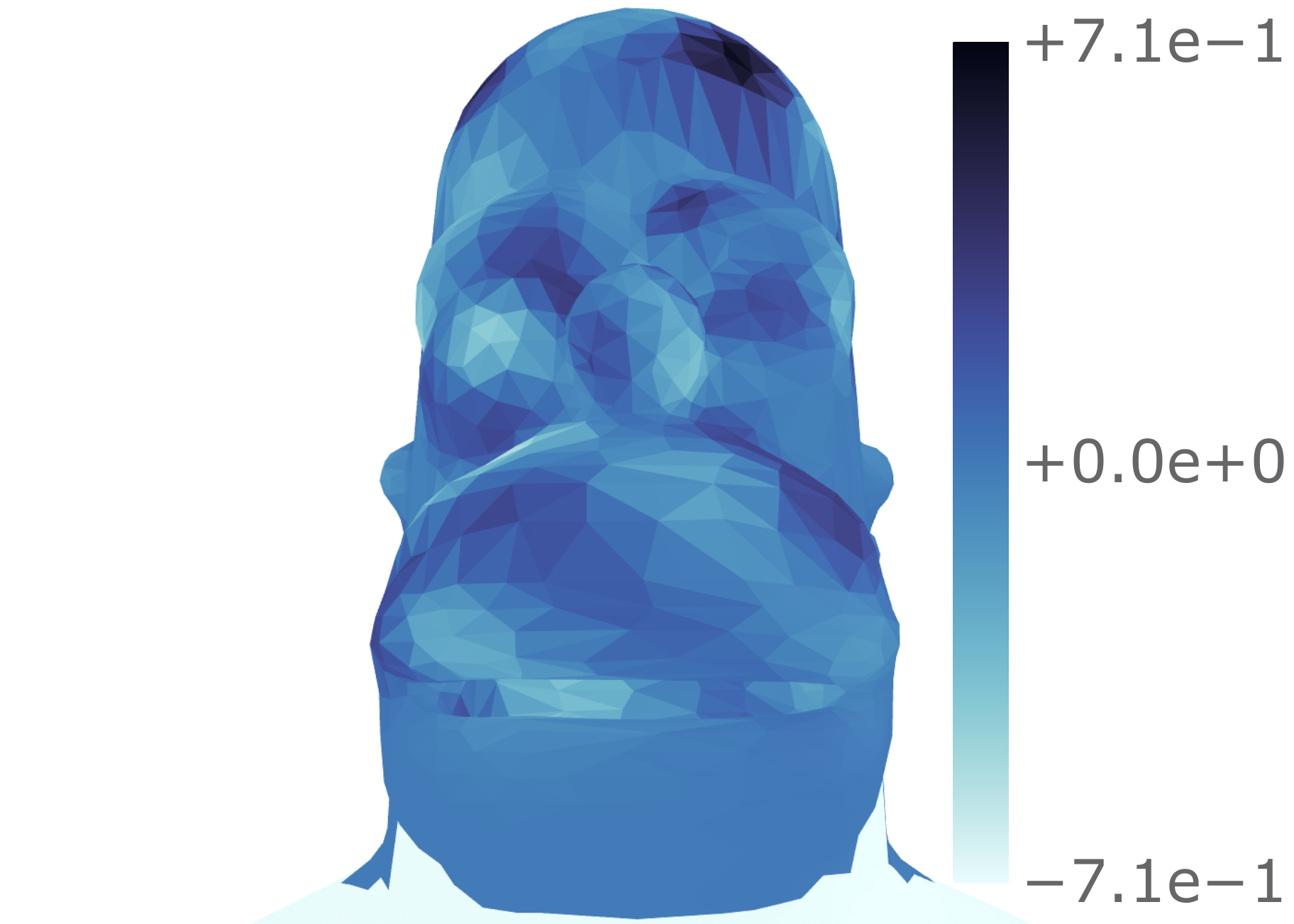}}
	\newline
	\subfloat[\(\mesh{\Psi^{4j}}\)]
	{\includegraphics[trim={101 0 3 3},clip,width=.33\columnwidth]{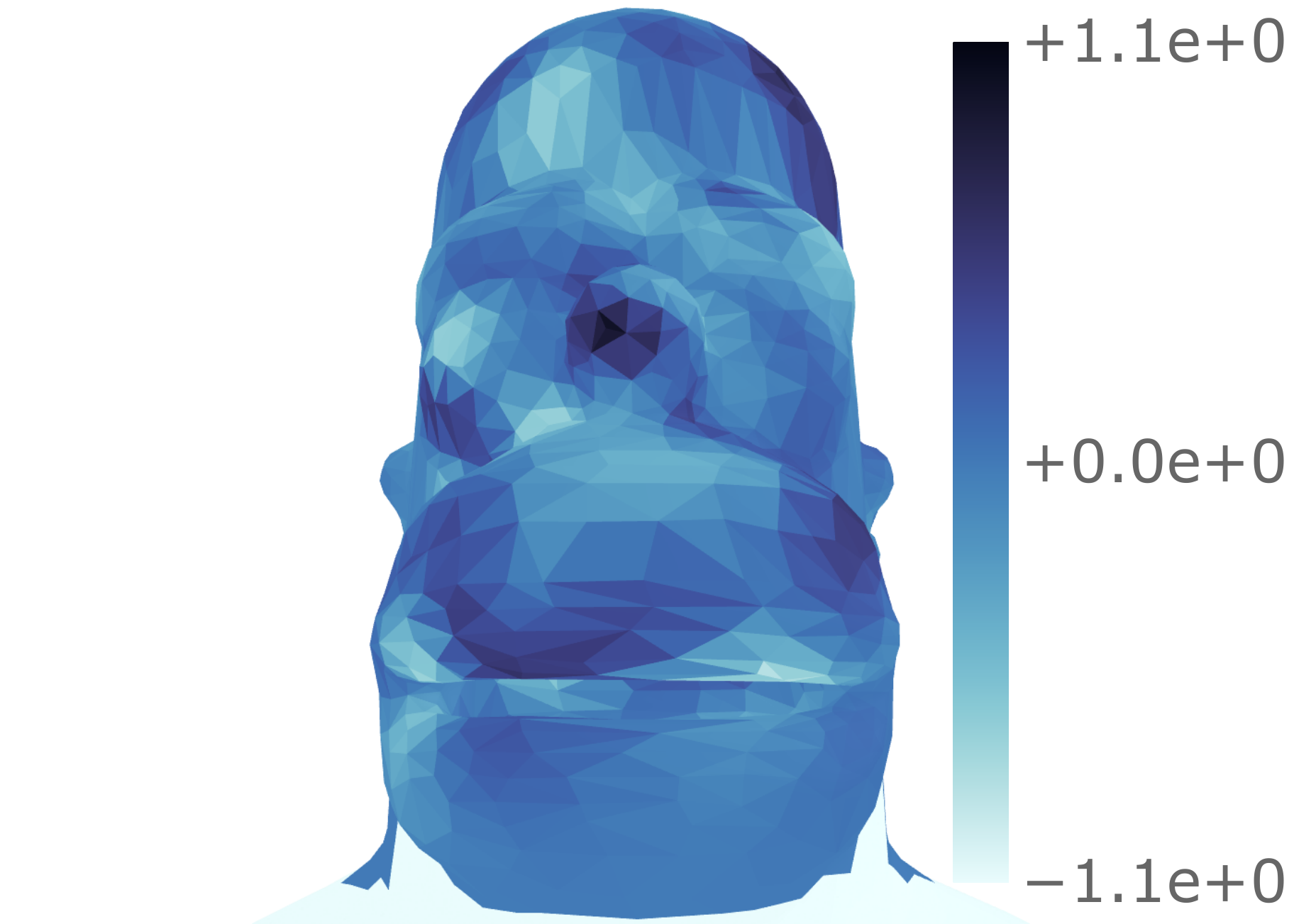}}
	\hfill
	\subfloat[\(\mesh{\Psi^{5j}}\)]
	{\includegraphics[trim={101 0 3 3},clip,width=.33\columnwidth]{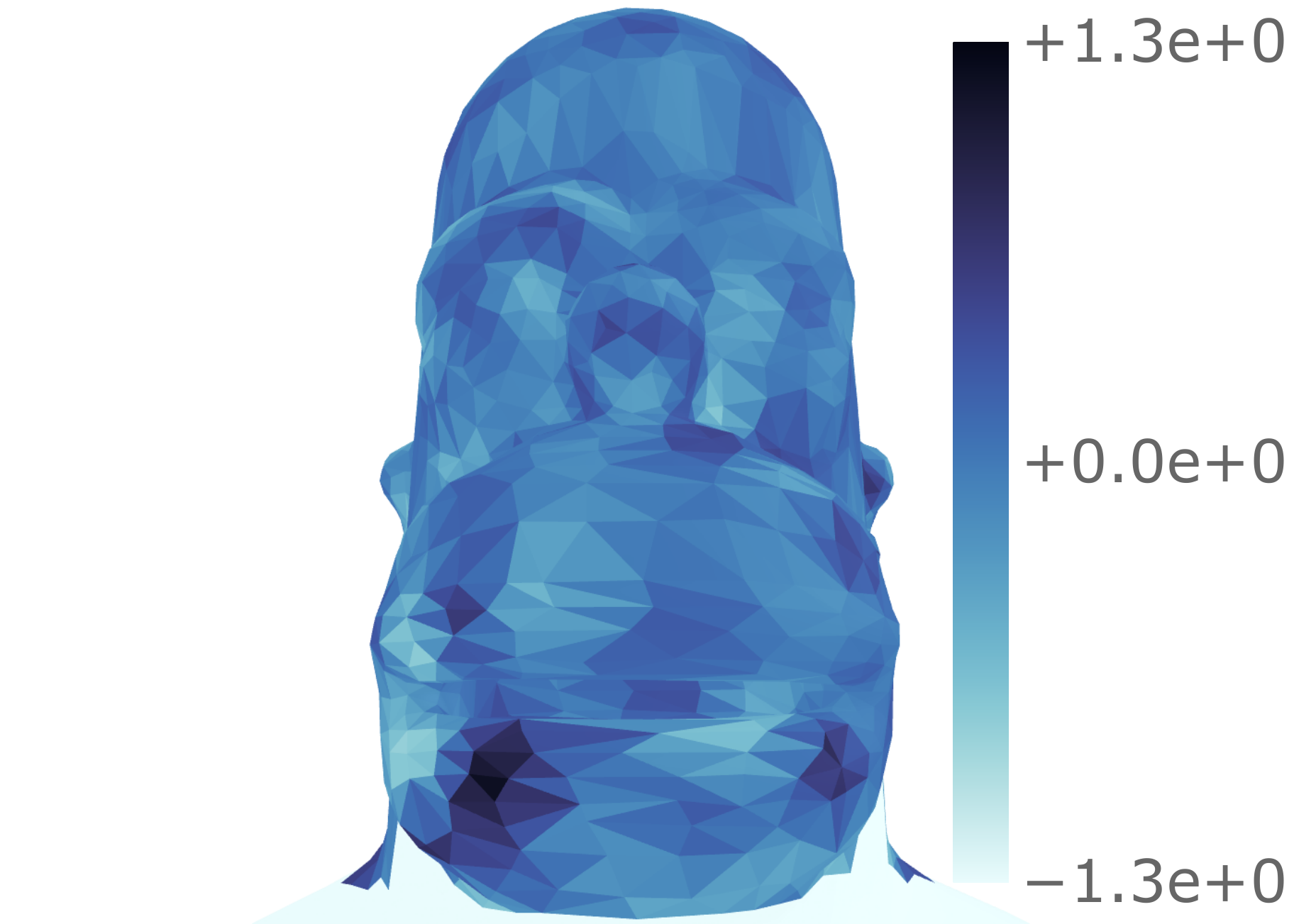}}
	\hfill
	\subfloat[\(\mesh{\Psi^{6j}}\)]
	{\includegraphics[trim={101 0 3 3},clip,width=.33\columnwidth]{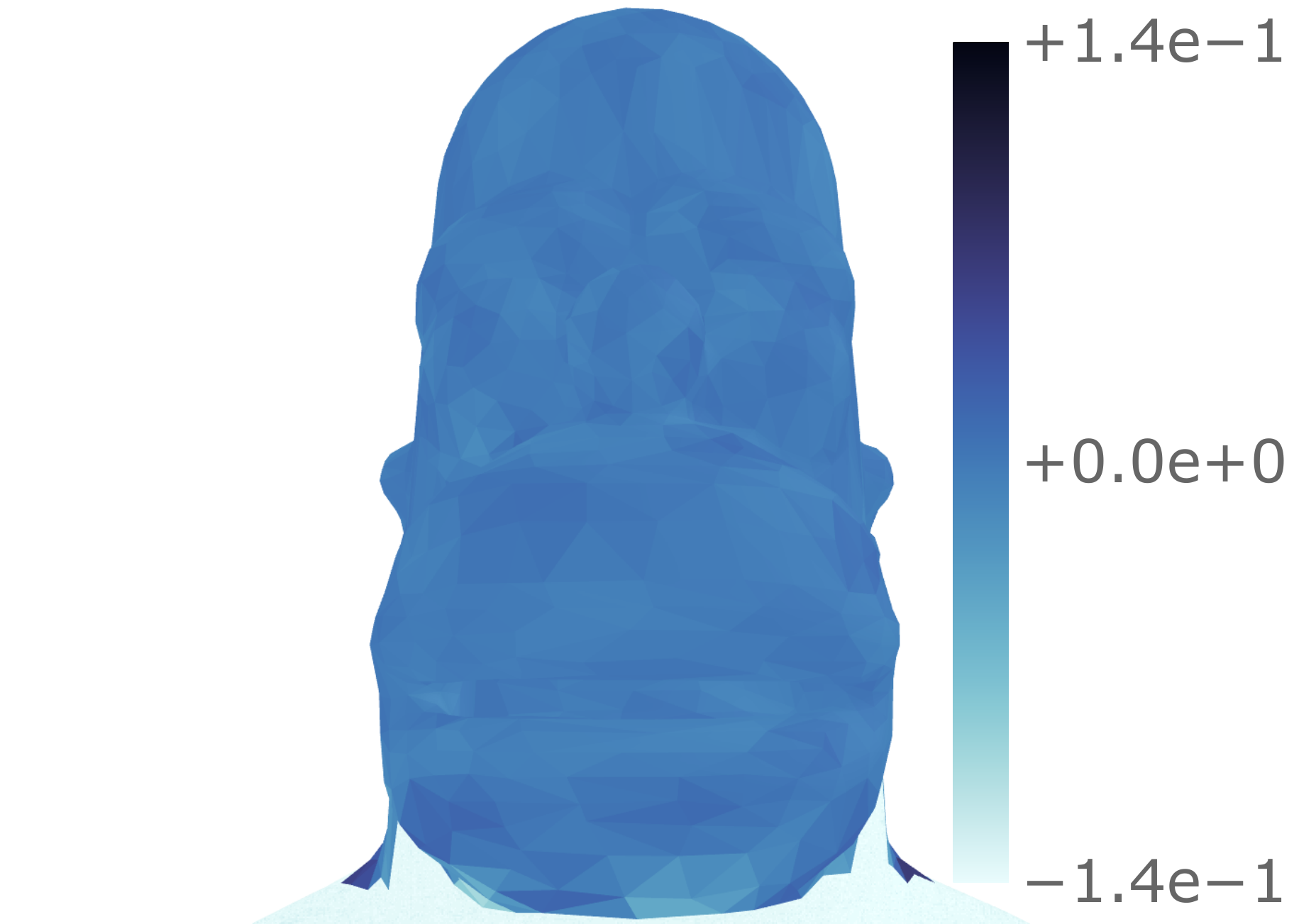}}
	\caption{
		The scaling function and the wavelets for scales \(j \in \set{2, 3, 4, 5, 6}\) for the Homer head region shown left-to-right, top-to-bottom.
		The wavelets are constructed through a tiling of the Slepian line using scale-discretised functions, with parameters \(\lambda=3\), \(J_{0}=2\), and \(\imax=\num{1275}\) basis functions.
		Whilst the wavelets are defined on the vertices, the values have been averaged onto the faces for the plot.
	}\label{fig:wavelets}
\end{figure}

\begin{figure}
    \hspace{20mm}
	\includegraphics[trim={156 8 21 6},clip,width=.62\columnwidth]{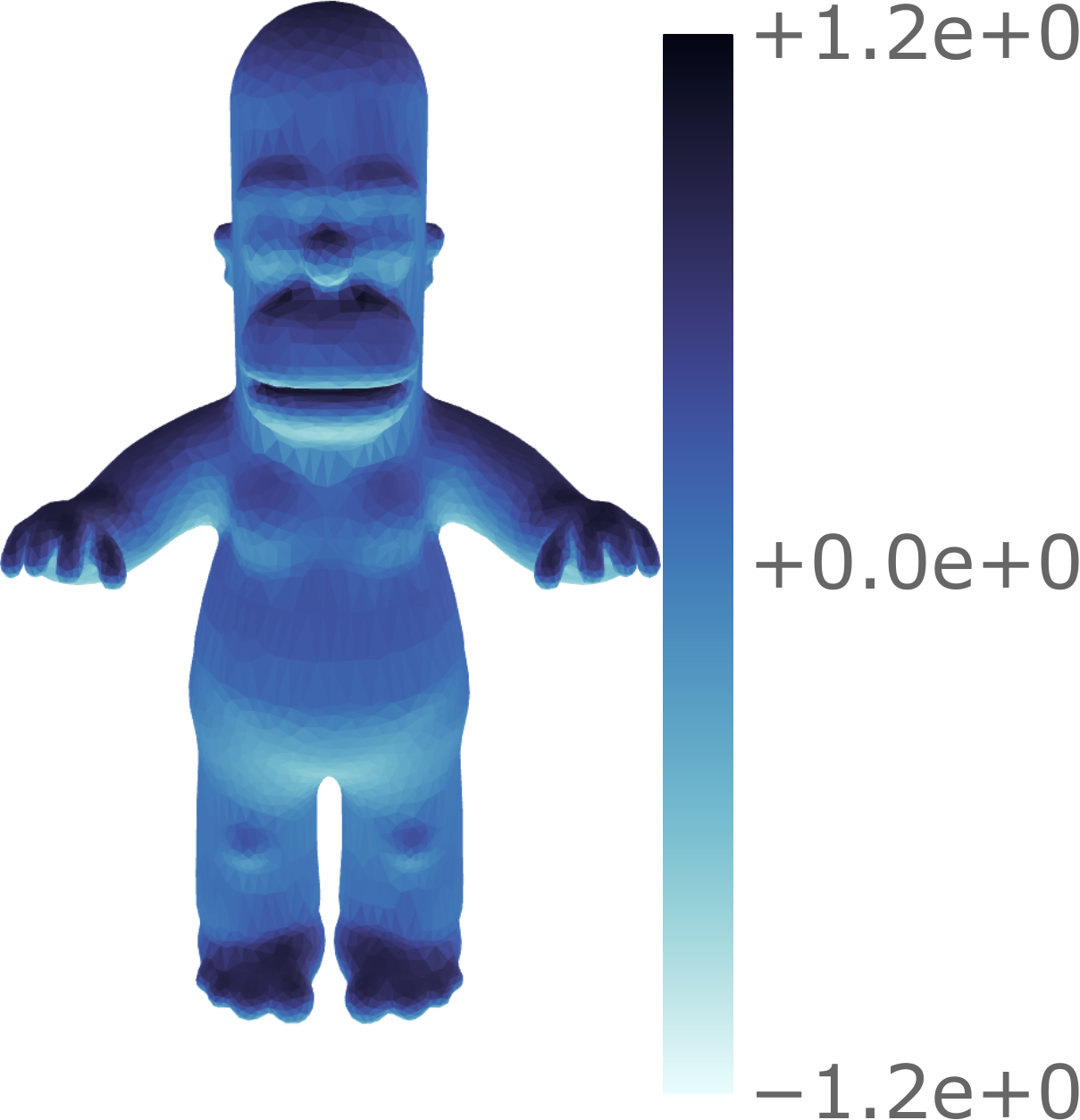}
	\caption{
		The \(z\)-component of the per-vertex normals of the Homer mesh.
		Whilst the field is defined on the vertices, the values have been averaged onto the faces for the plot.
	}\label{fig:homer_data}
\end{figure}

\begin{figure}
	\centering
	\subfloat[\(\mesh{W^{\Phi}}\)]
	{\includegraphics[trim={101 0 3 3},clip,width=.33\columnwidth]{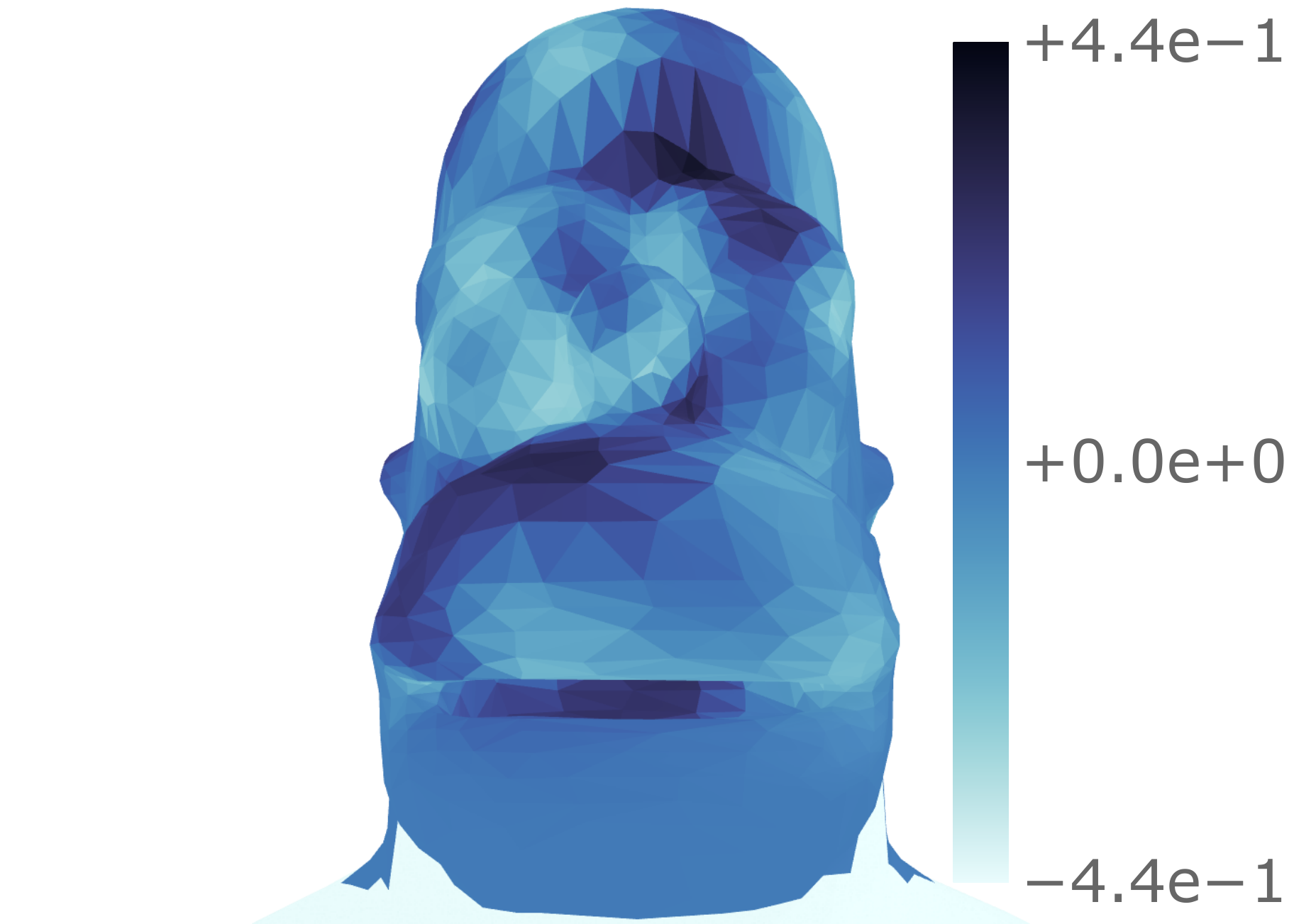}}
	\hfill
	\subfloat[\(\mesh{W^{\Psi^{2j}}}\)]
	{\includegraphics[trim={101 0 3 3},clip,width=.33\columnwidth]{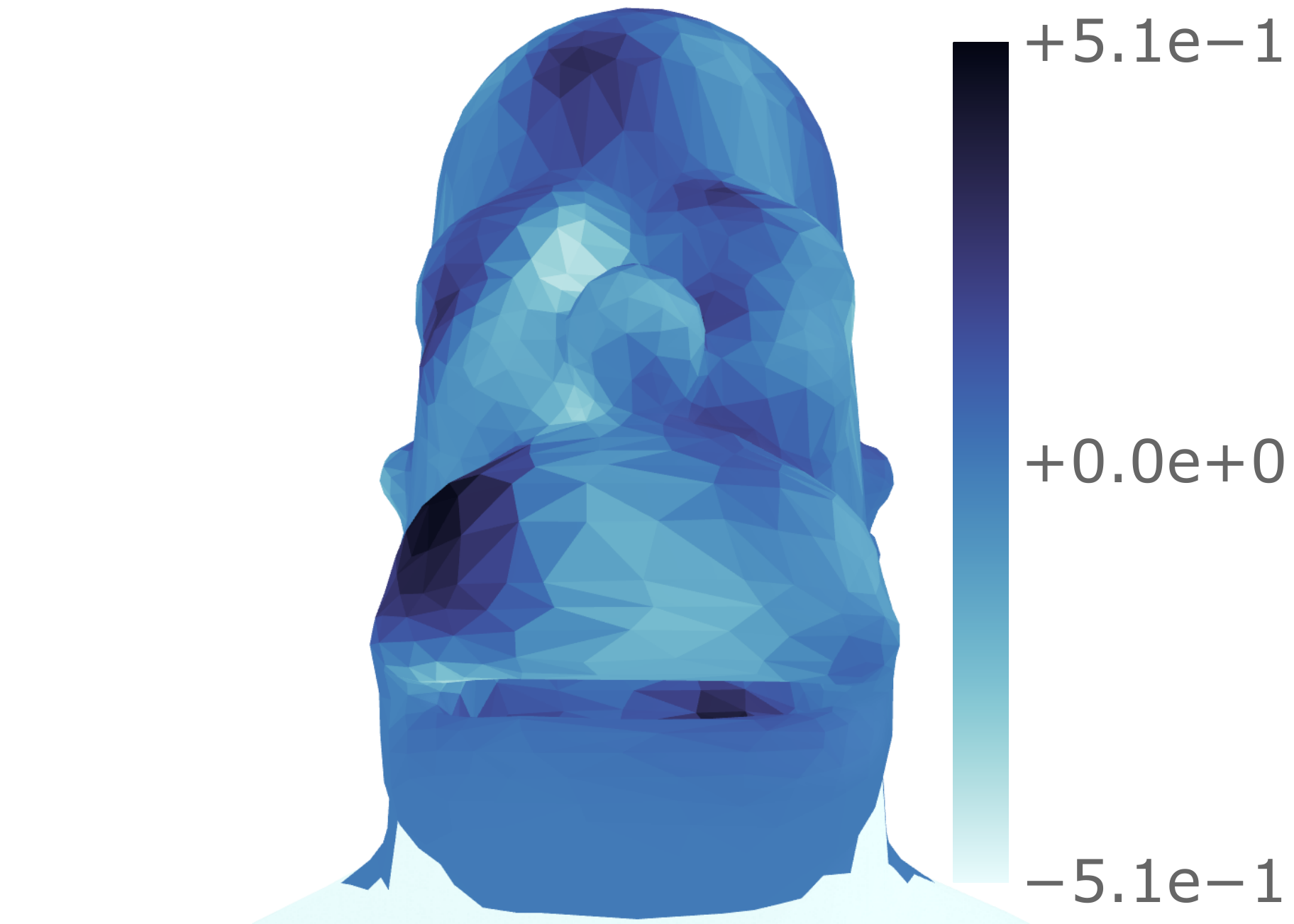}}
	\hfill
	\subfloat[\(\mesh{W^{\Psi^{3j}}}\)]
	{\includegraphics[trim={101 0 3 3},clip,width=.33\columnwidth]{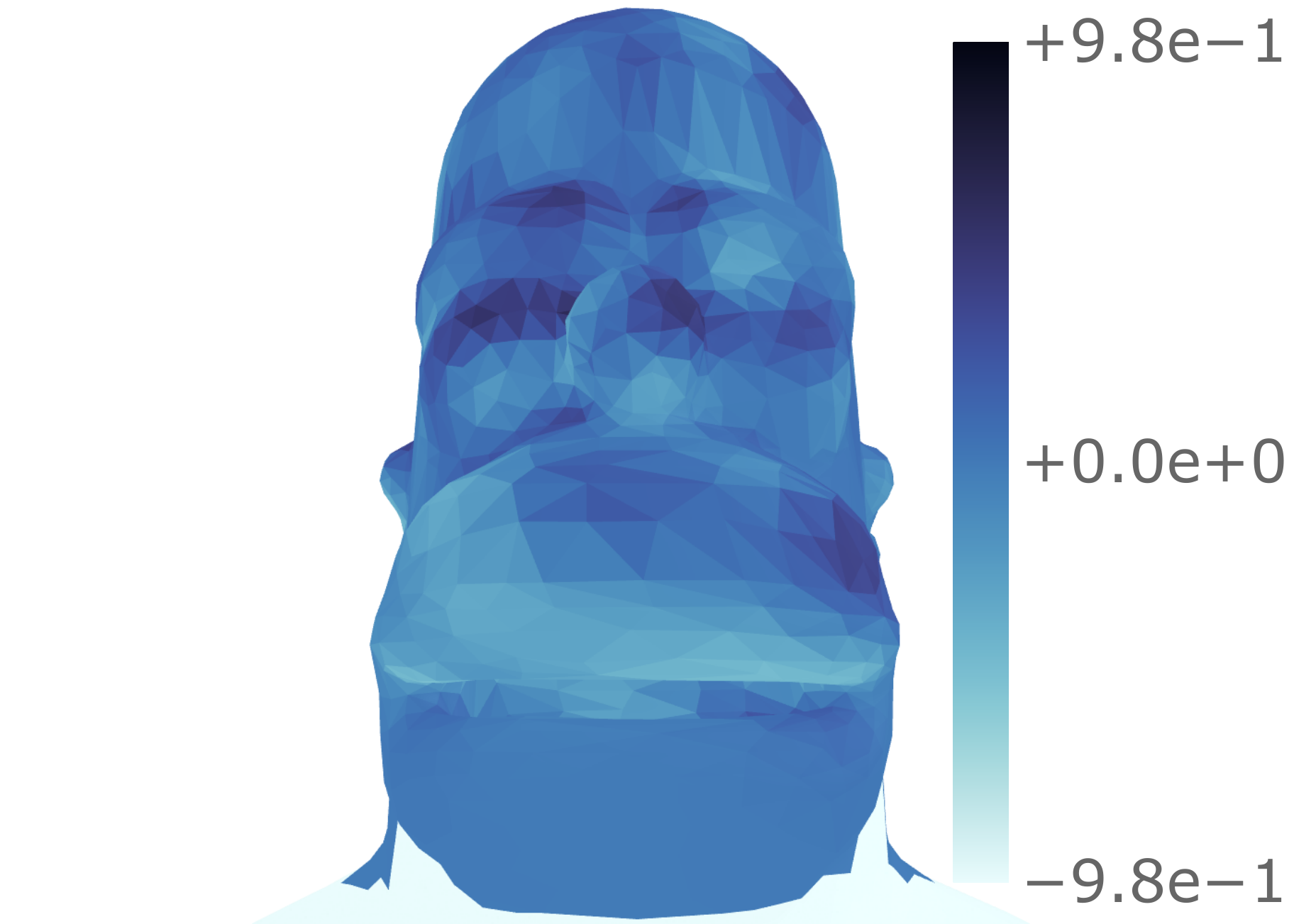}}
	\newline
	\subfloat[\(\mesh{W^{\Psi^{4j}}}\)]
	{\includegraphics[trim={101 0 3 3},clip,width=.33\columnwidth]{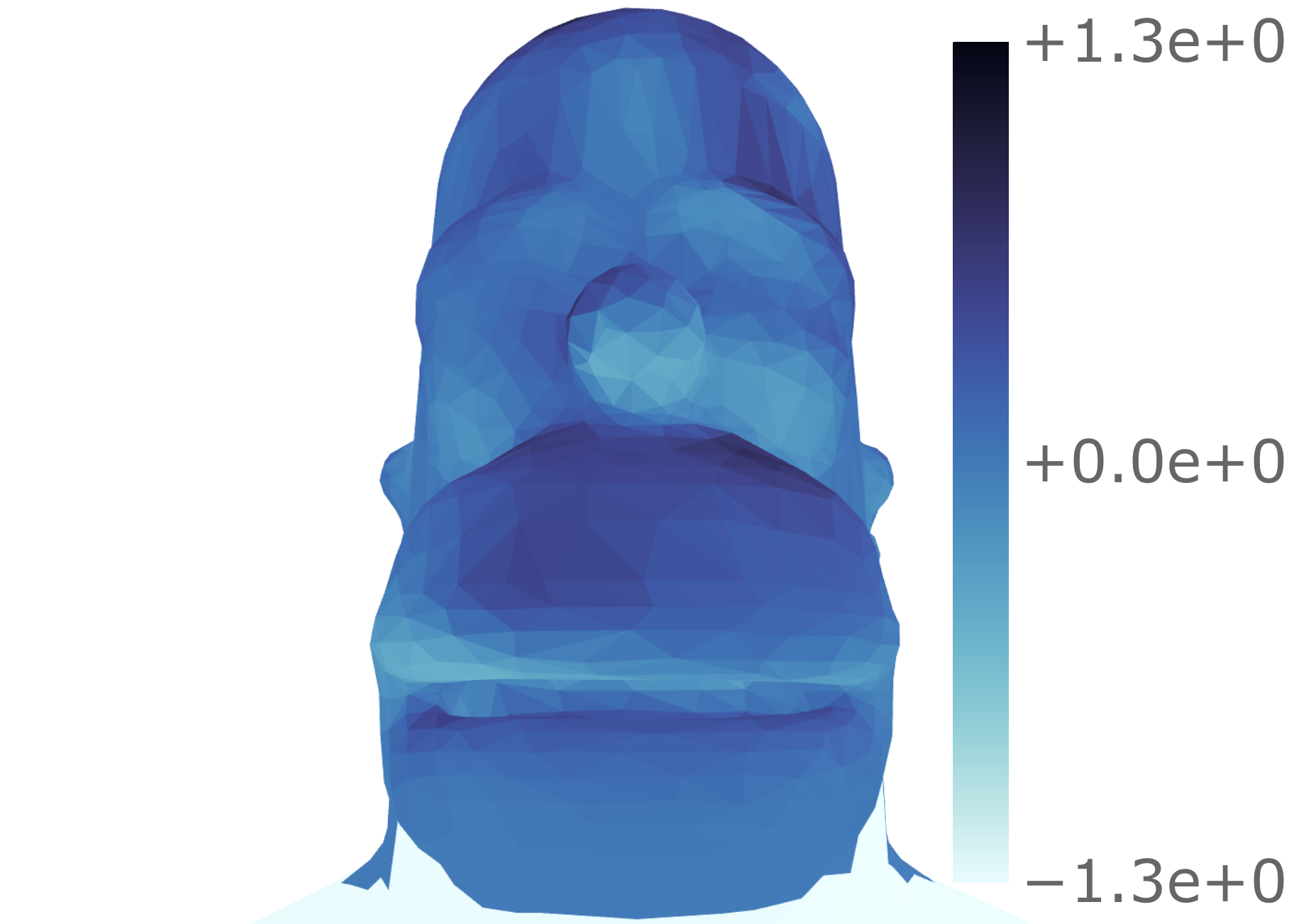}}
	\hfill
	\subfloat[\(\mesh{W^{\Psi^{5j}}}\)]
	{\includegraphics[trim={101 0 3 3},clip,width=.33\columnwidth]{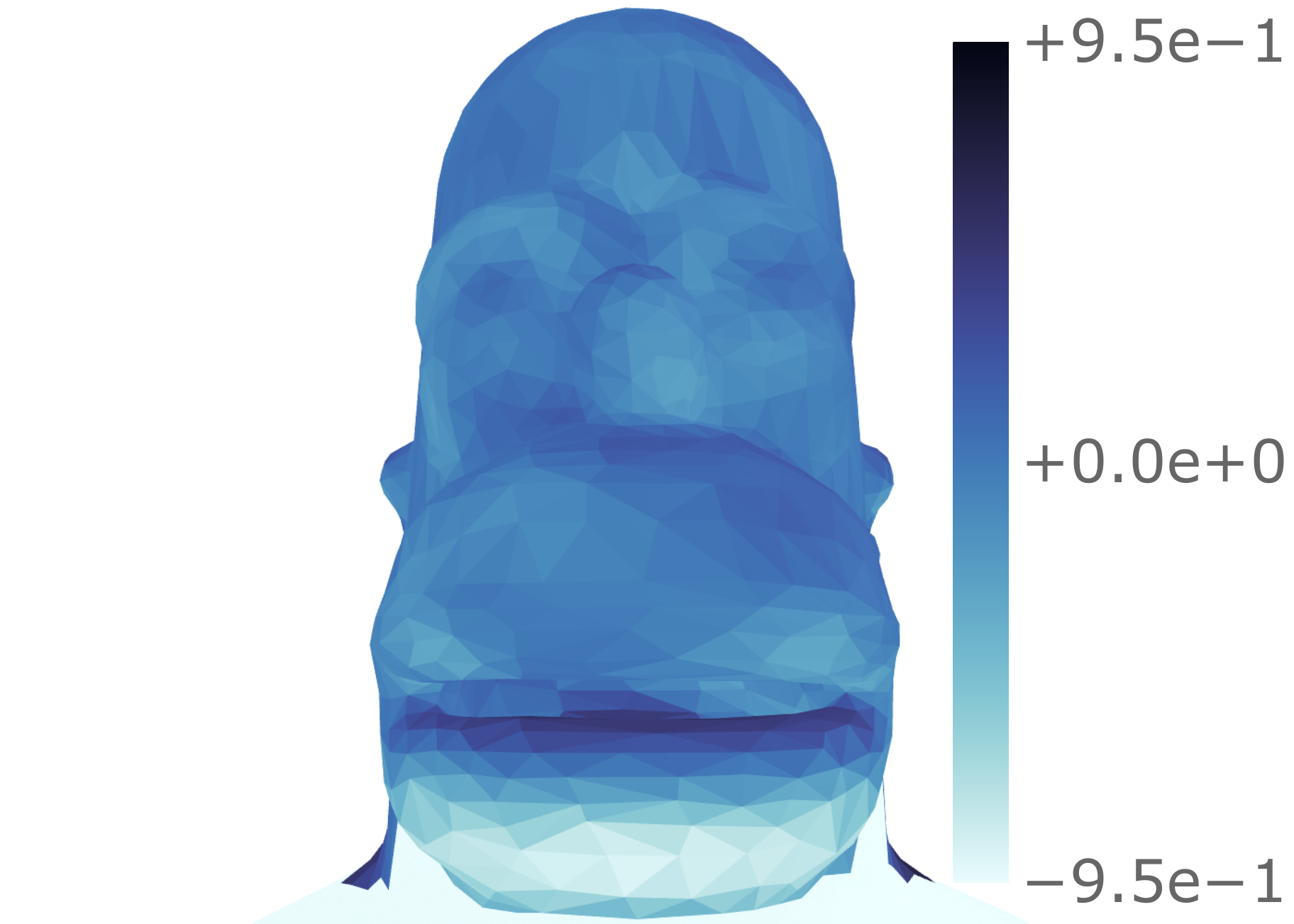}}
	\hfill
	\subfloat[\(\mesh{W^{\Psi^{6j}}}\)]
	{\includegraphics[trim={101 0 3 3},clip,width=.33\columnwidth]{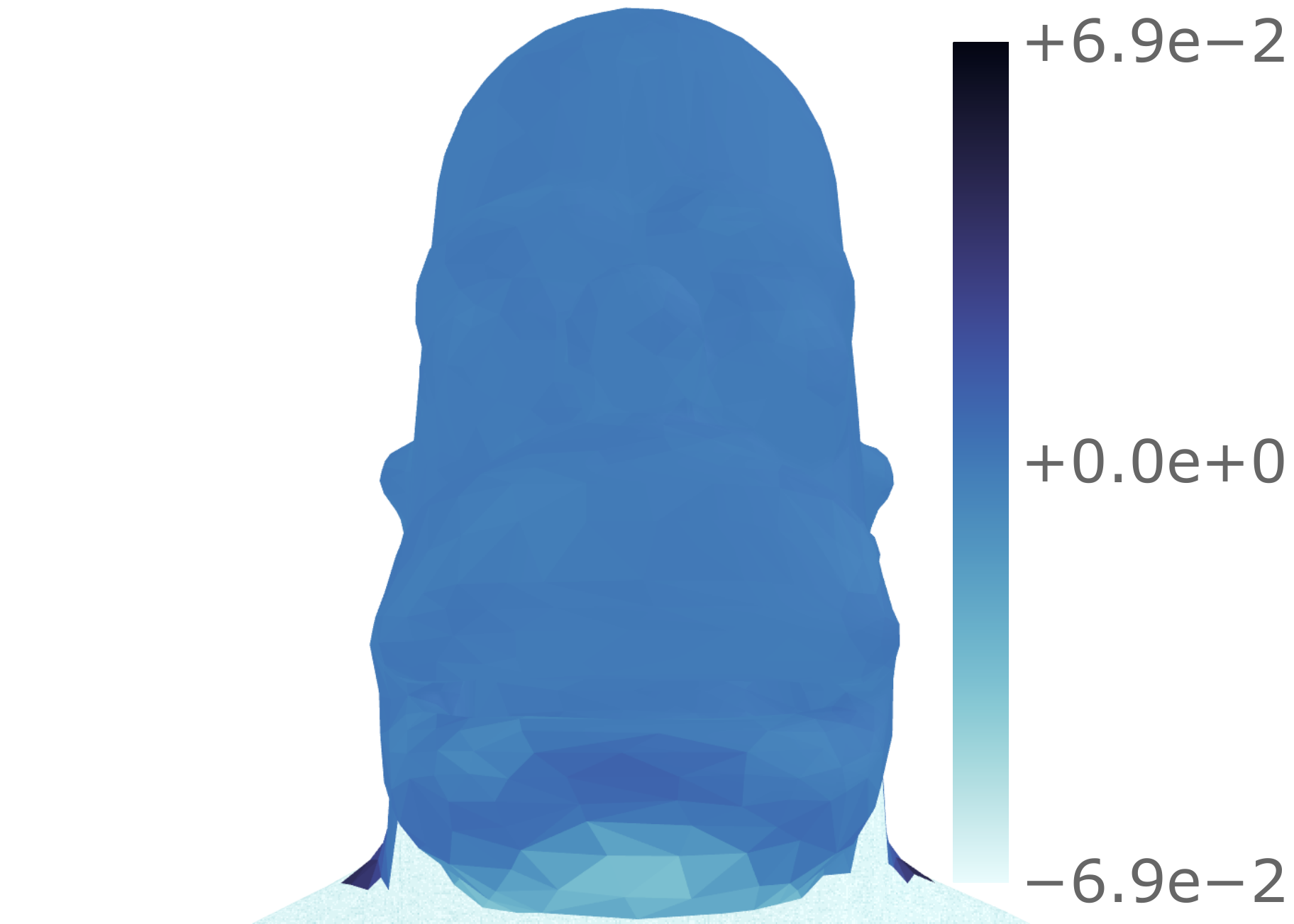}}
	\caption{
		The scale-discretised wavelet transform of the per-vertex normals of the Homer head region for parameters \(\lambda=3\), \(J_{0}=2\), and \(\imax=\num{1275}\) basis functions; \ie{} with the wavelets shown in \cref{fig:wavelets}.
		Spatially localised, scale-dependent features of the bandlimited signal may be extracted by the wavelet coefficients given by the wavelet transform.
		The scaling coefficients are given in the top left plot, while the wavelet coefficients at scales \(j \in \set{2, 3, 4, 5, 6}\) are shown left-to-right, top-to-bottom.
		Whilst the wavelet coefficients are defined on the vertices, the values have been averaged onto the faces for the plot.
	}\label{fig:wavelet_coefficients}
\end{figure}

The Slepian scaling function and wavelets defined in \cref{sec:slepian_scale_discretised_wavelets} are built on a tiling of the Slepian line with parameters \(\lambda=3\) and \(J_{0}=2\).
This tiling is shown in \cref{fig:tiling} for \(\imax=\num{1275}\) basis functions of the Homer mesh, where the Shannon number \(N=329\) is highlighted.
Hence, for this region, the scaling function and wavelets for scales \(j \in \set{2, 3, 4, 5, 6}\) are the only non-zero functions.
These wavelets are presented in \cref{fig:wavelets}, in which a similar pattern to the Slepian functions is observed whereby the scaling function is more concentrated than the \(j=6\) wavelet.
To perform a scale-discretised wavelet transform, one requires a signal on the mesh.
\cref{fig:homer_data} presents such a signal, the \(z\)-component of the per-vertex normals of the Homer mesh.
With some data to hand, the scaling and wavelet coefficients of the Homer head region are given in \cref{fig:wavelet_coefficients}.

\subsection{Wavelet Denoising}\label{sec:wavelet_denoising}

Wavelets are used in a variety of contexts in signal processing, one common use case is for denoising a signal.
Localised features in the data can be extracted to different wavelet scales, and hence the desired parts of the signal can be preserved whilst isolating the noise.
To showcase Slepian wavelets, white noise is added to the signal in \cref{fig:homer_data}.
A straightforward hard-thresholding denoising procedure follows.

Consider a signal localised in the region \(R\) in the presence of noise
\begin{equation}\label{eq:noised_signal}
	\mesh{z}
	= \mesh{s} + \mesh{n},
\end{equation}
where the signal and noise are represented by \(\mesh{s}\) and \(\mesh{n}\) respectively.
The power spectrum of noise in Slepian space is
\begin{equation}
	\expval*{\slepian{n} \conj{\slepian[']{n}}}
	= \sigma^{2} \delta_{pp'}.
\end{equation}
A denoised version of \(z\) is desired, denoted by \(d \in \hilbert{R}\), with a large \(\snr{d}\) such that \(d\) extracts the informative signal \(s\).
The scaling function and wavelets are treated equivalently in the Slepian setting.
The scaling function in the Slepian setting is not a low-frequency representation of the signal due to the localisation of the Slepian functions.

Since the wavelet transform is linear, the individual elements sum to give the wavelet/scaling coefficients of \cref{eq:noised_signal}
\begin{equation}
	\mesh{Z^{\varphi}}
	= \mesh{S^{\varphi}} + \mesh{N^{\varphi}},
\end{equation}
where the wavelet coefficients are denoted by capital letters, \ie{} \(\mesh{Z^{\varphi}} = \mesh{\convolution{\varphi}{z}}\).
In wavelet space the noise is
\begin{equation}
	\variance{\mesh{N^{\varphi}}}
	= \sigma^{2} \slepianSum \abs{\slepian{\varphi}}^{2} \abs{\mesh{\slepian{S}}}^{2}
	\equiv {\mesh{\sigma^{\varphi}}}^{2},
\end{equation}
where \(\sigma^{\varphi}\) represents the standard deviation of the noise in wavelet space.
To denoise the signal one may hard-threshold the scaling/wavelet coefficients with a threshold \(T\) proportional to the standard deviation of the noise.
Hence, the denoised wavelet coefficients \(\mesh{D^{\varphi}}\) are
\begin{equation}
	\mesh{D^{\varphi}} =
	\begin{cases}
		0,
		 & \mesh{Z^{\varphi}} < \mesh{T^{\varphi}},    \\
		\mesh{Z^{\varphi}},
		 & \mesh{Z^{\varphi}} \geq \mesh{T^{\varphi}},
	\end{cases}
\end{equation}
where
\begin{equation}
	\mesh{T^{\varphi}}
	= N_{\sigma}\mesh{\sigma^{\varphi}},
\end{equation}
with \(N_{\sigma} \in \realPosParam{}\).
Reconstruction of the signal \(d\) follows by an inverse wavelet transform with these thresholded wavelet coefficients.
This procedure merely demonstrates a practical use case of Slepian wavelets, more sophisticated denoising formalisms can be developed.

\begin{figure*}
	\centering
	\subfloat[Initial Data]
	{\includegraphics[trim={101 0 3 3},clip,width=.5\columnwidth]{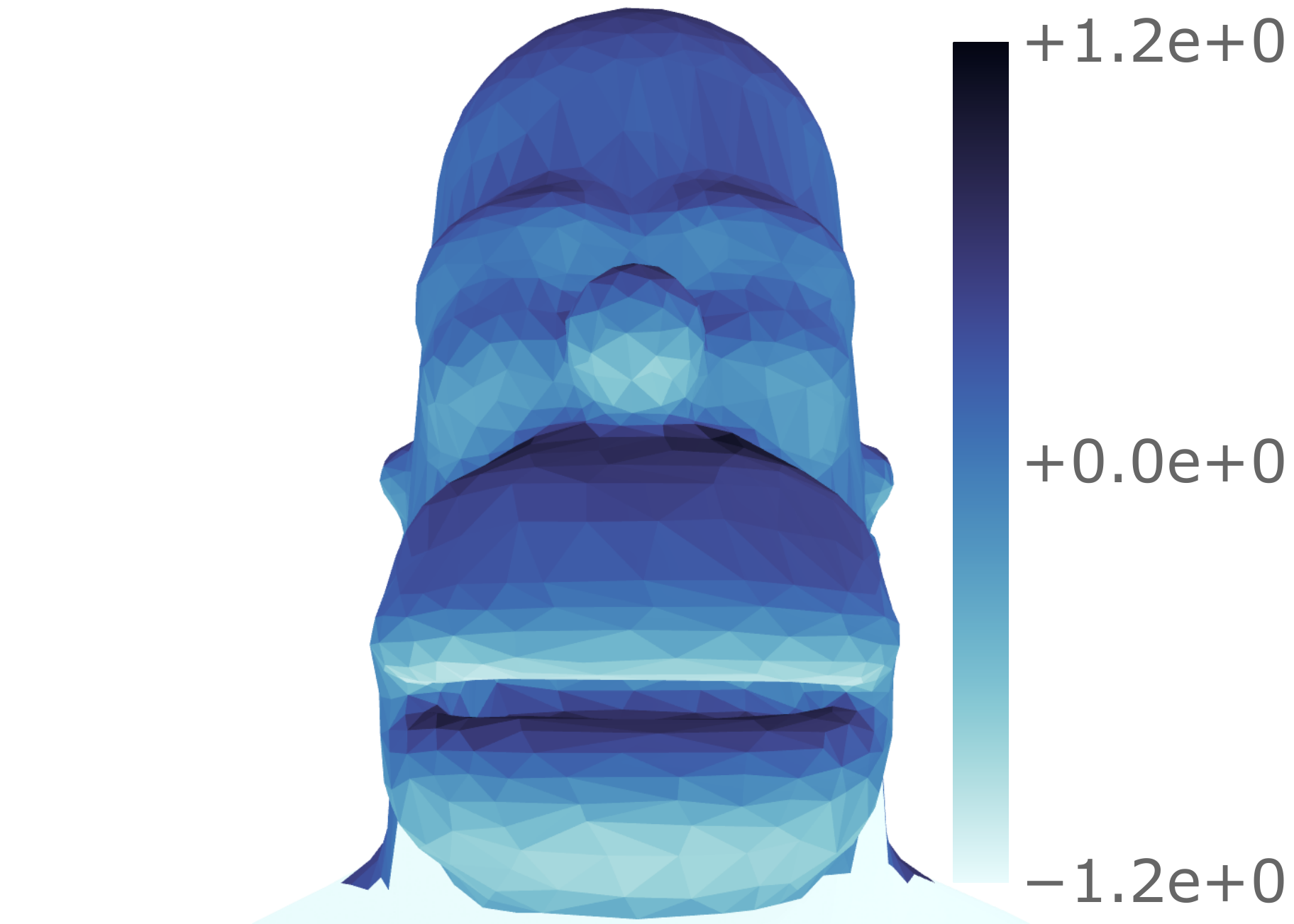}}
	\subfloat[Noisy Data \newline
		\(\snr{z} = \SI{0.32}{\dB}\)]
	{\includegraphics[trim={101 0 3 3},clip,width=.5\columnwidth]{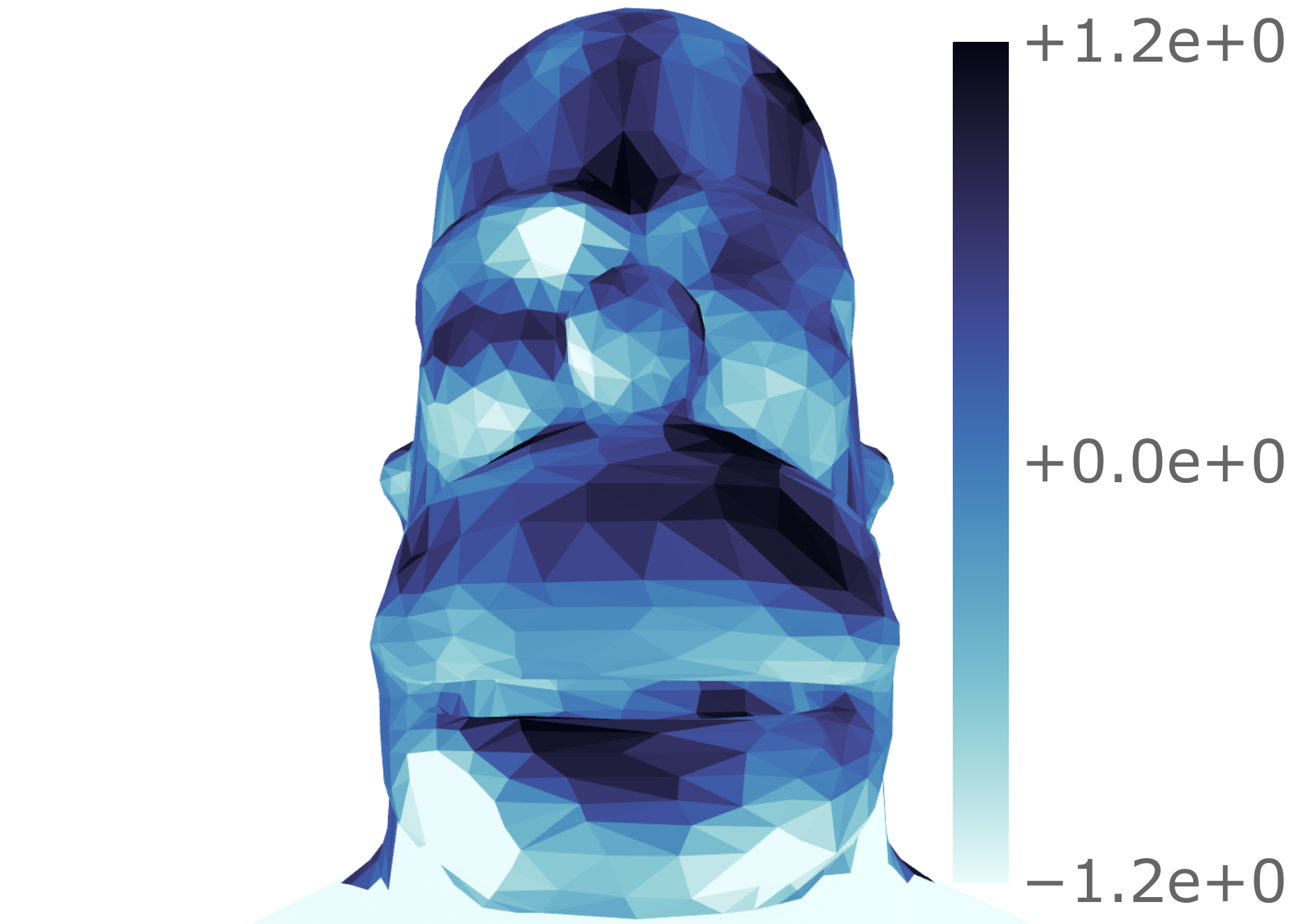}}
	\subfloat[Denoised \(N_{\sigma}=2\) \newline
		\(\snr{d} = \SI{2.25}{\dB}\)]
	{\includegraphics[trim={101 0 3 3},clip,width=.5\columnwidth]{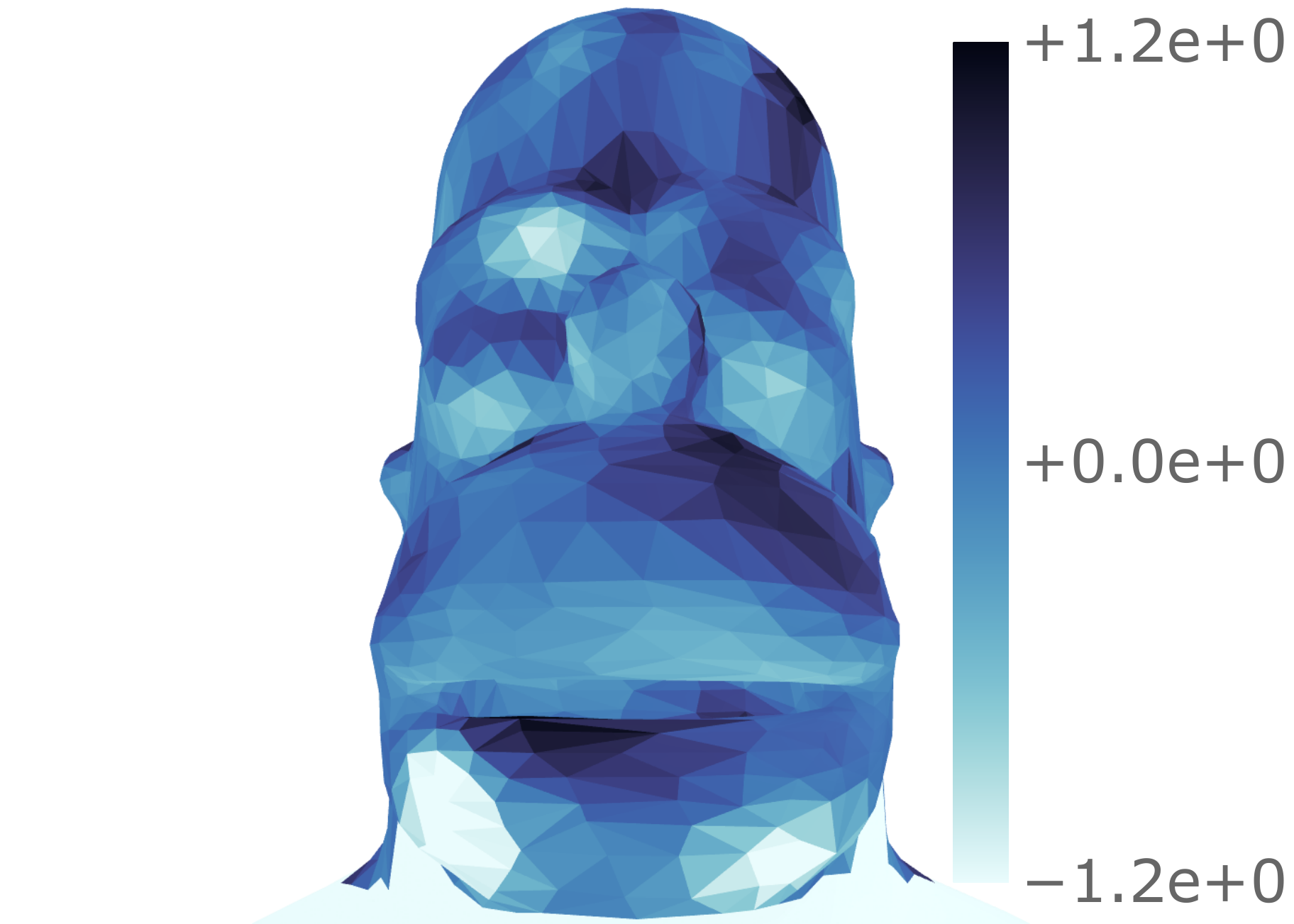}}
	\caption{
		Panel (a) shows the data in the region \(R\) constructed from the Slepian coefficients of the per-vertex normals (\cf{} \cref{fig:homer_data}) --- where the field value outside the region is set to negative infinity for illustrative purposes.
		Gaussian white noise is added to the signal in the Homer head region with a signal-to-noise ratio of \(\SI{0.32}{\dB}\), shown in panel (b).
		The scaling and wavelet coefficients of the noisy signal are calculated and are then hard-thresholded with \(N_{\sigma}=2\).
		The corresponding denoised plot is shown in panel (c), where the signal-to-noise ratio is boosted by \(\SI{1.93}{\dB}\) to \(\SI{2.25}{\dB}\).
		Whilst the signal values are defined on the vertices, they have been averaged onto the faces for the plot.
	}\label{fig:denoising}
\end{figure*}

\begin{figure*}
	\centering
	\subfloat[Cheetah
		(\(\num{2002}\, \mathcal{V}\))]
	{\includegraphics[trim={137 1 3 7},clip,width=.3\columnwidth]{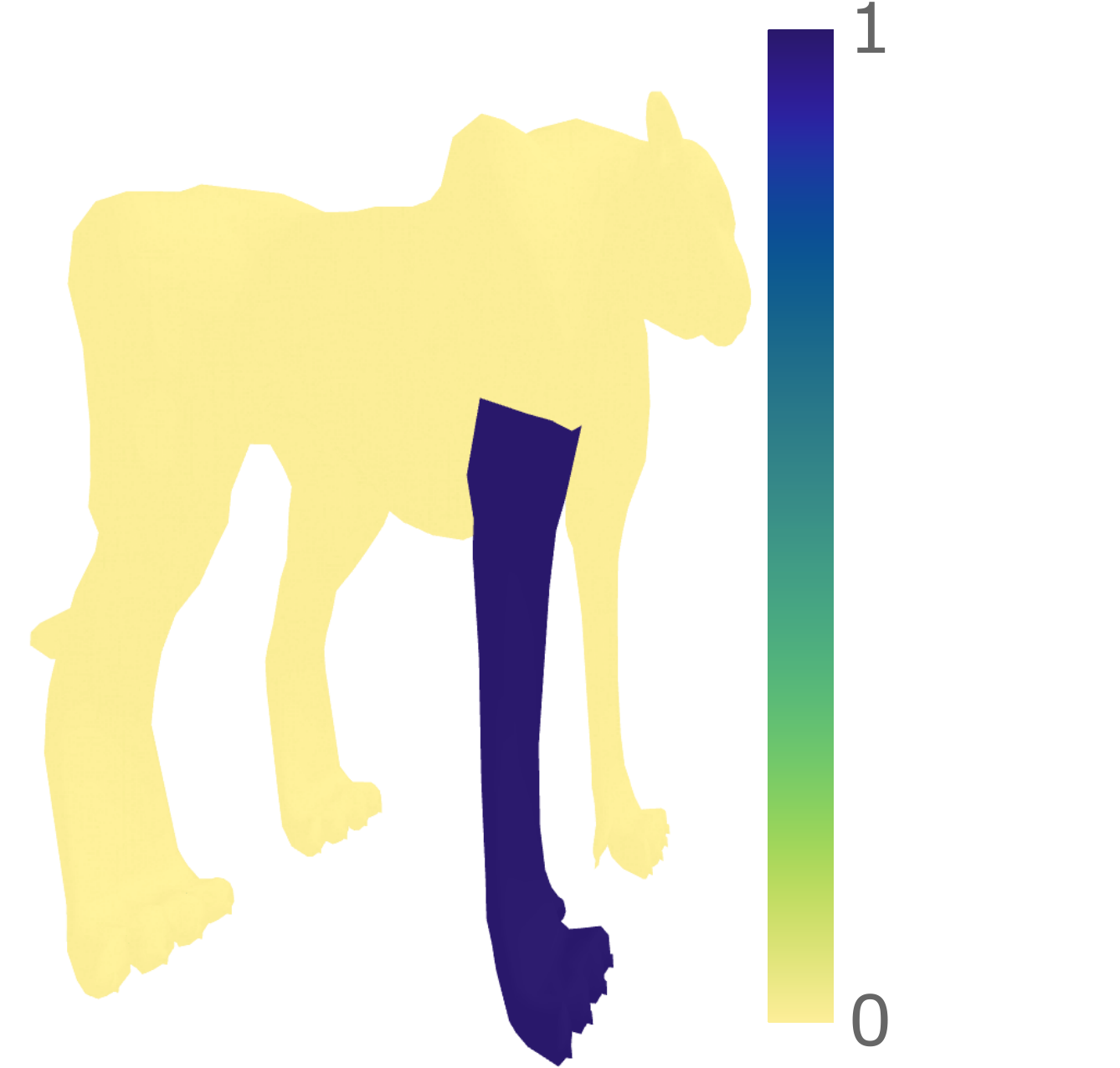}}
	\hfill
	\subfloat[Dragon
		(\(\num{1166}\, \mathcal{V}\))]
	{\includegraphics[trim={75 8 3 7},clip,width=.36\columnwidth]{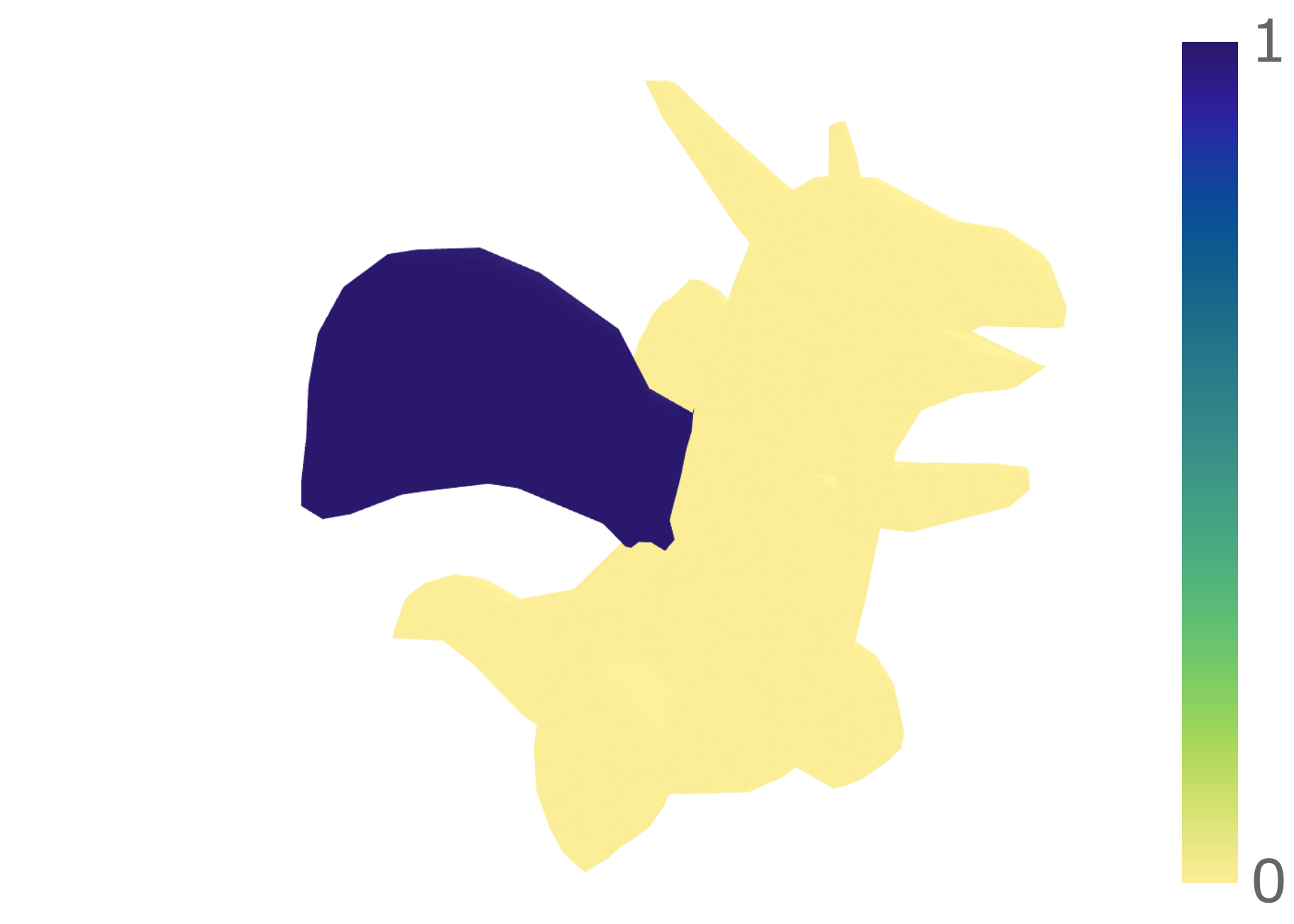}}
	\hfill
	\subfloat[Bird
		(\(\num{699}\, \mathcal{V}\))]
	{\includegraphics[trim={7 8 3 7},clip,width=.42\columnwidth]{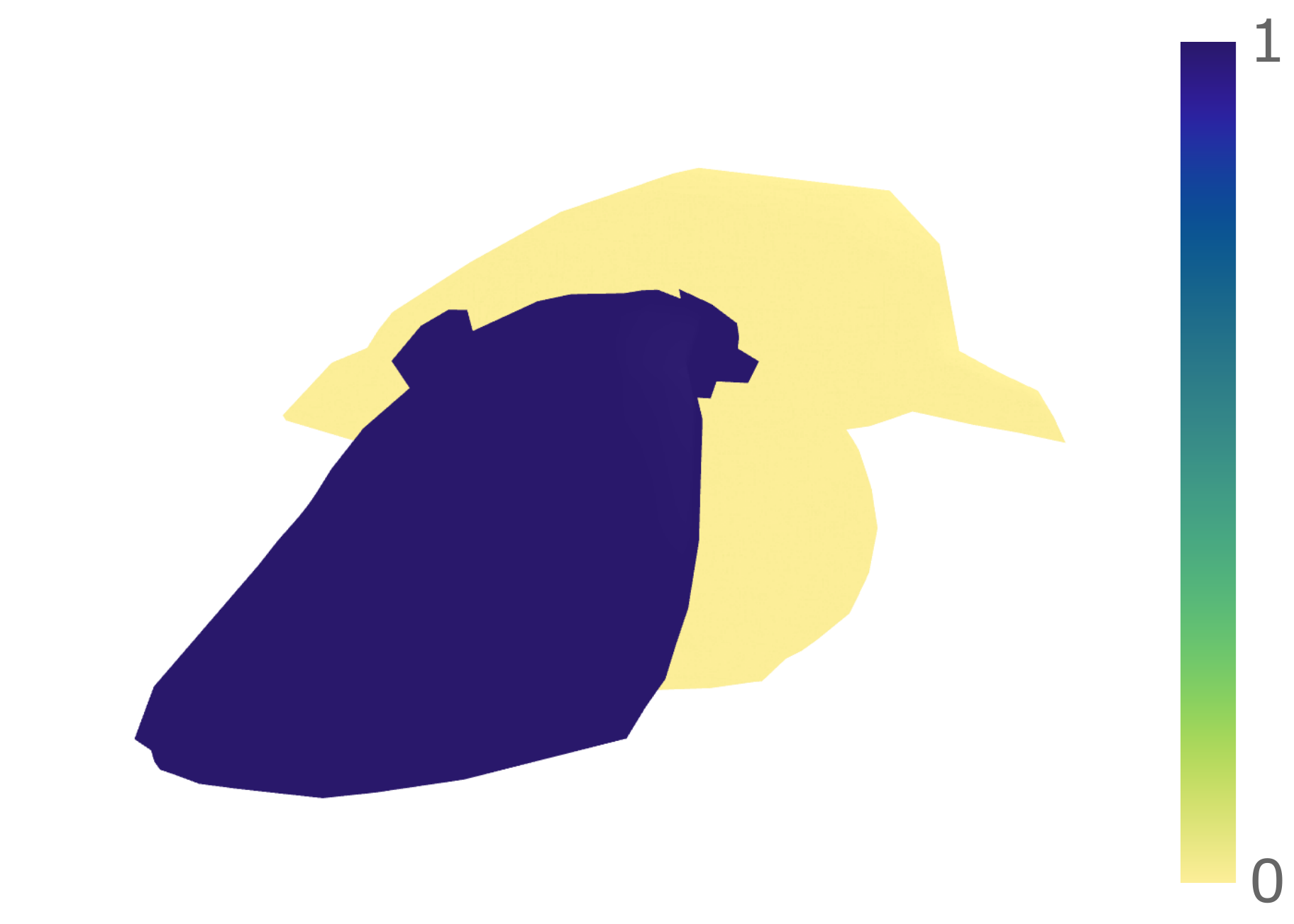}}
	\hfill
	\subfloat[Teapot
		(\(\num{726}\, \mathcal{V}\))]
	{\includegraphics[trim={3 8 3 7},clip,width=.42\columnwidth]{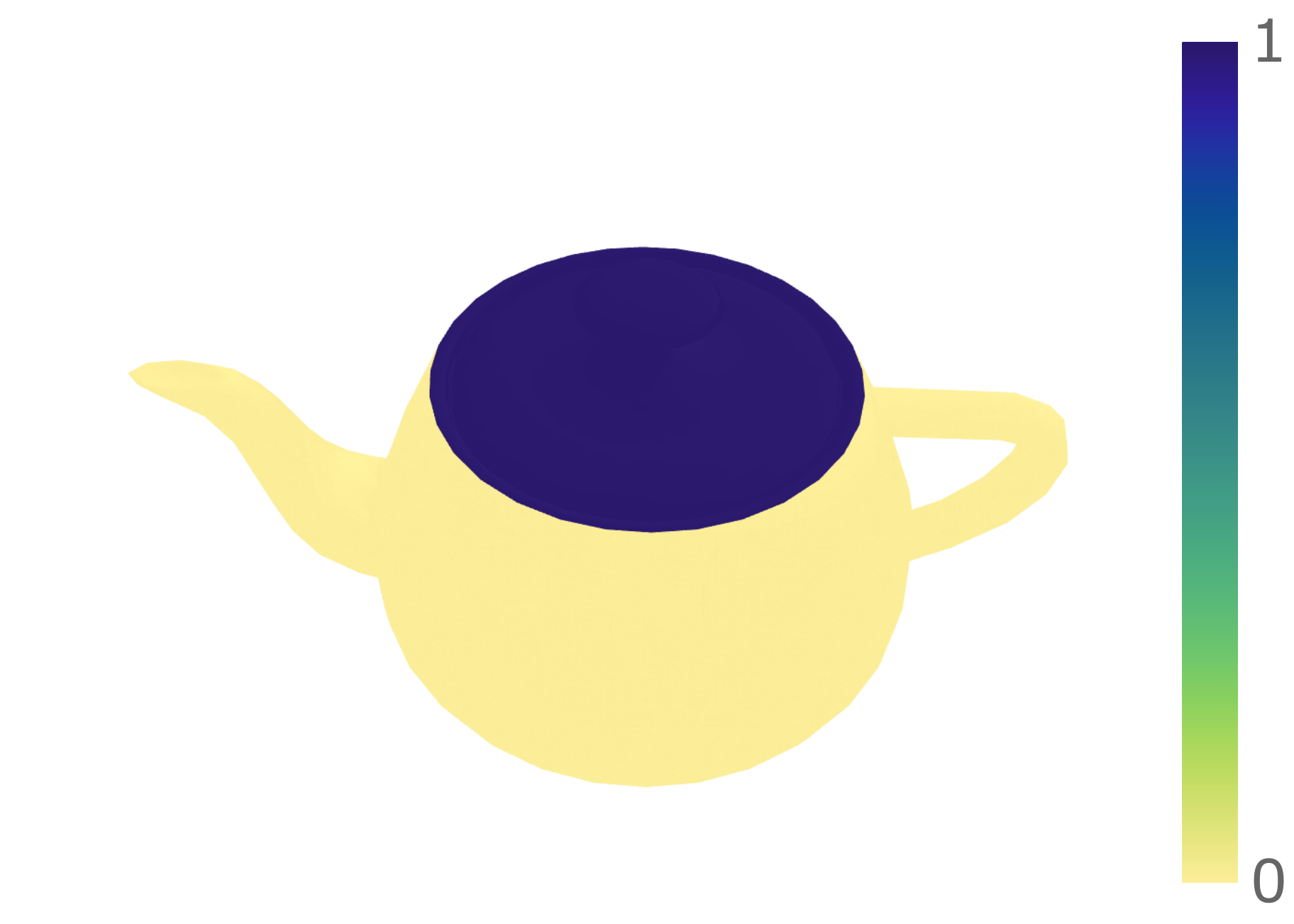}}
	\hfill
	\subfloat[Cube
		(\(\num{6146}\, \mathcal{V}\))]
	{\includegraphics[trim={62 1 3 7},clip,width=.37\columnwidth]{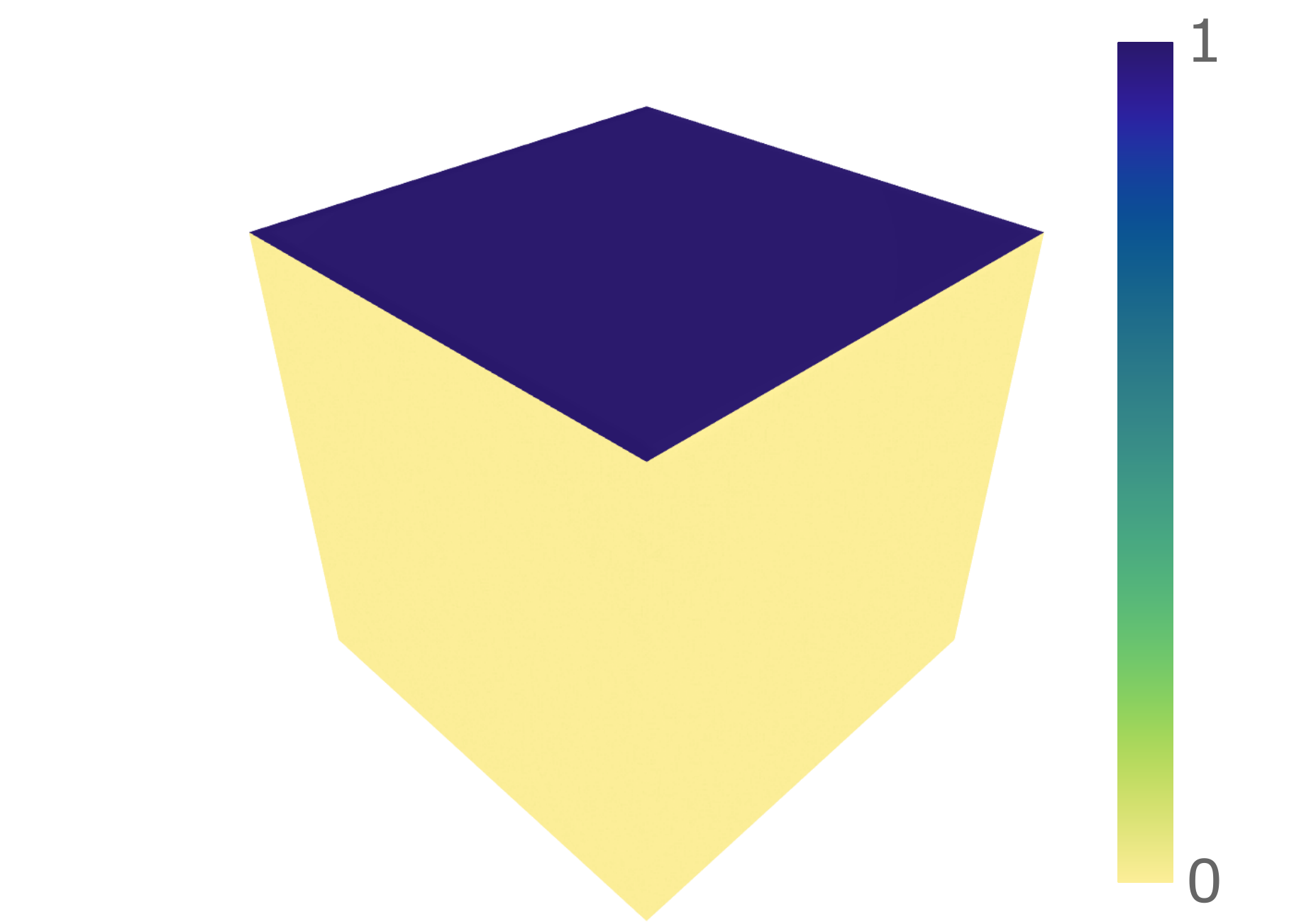}}
	\caption{
		The Slepian regions (in blue) and the corresponding number of vertices \(\vertices{}\) of some other meshes.
		The same denoising procedure as in \cref{fig:denoising} was performed for these alternative meshes, the results of which are shown in \cref{tab:denoising}.
	}\label{fig:other_meshes}
\end{figure*}

To perform the denoising, Gaussian white noise is added to the data in panel (a) of \cref{fig:denoising}.
Panel (b) presents this noised signal with an initial signal-to-noise ratio of \(\SI{0.32}{\dB}\).
The result of the denoising procedure described above for \(N_{\sigma}=2\) is shown in panel (c), where a signal-to-noise ratio of \(\SI{2.25}{\dB}\) represents a boost of \(\SI{1.93}{\dB}\).
The same denoising procedure was conducted for a series of other meshes, the Slepian regions of which are shown in \cref{fig:other_meshes}.
These meshes are comparable in size to the \(\num{5103}\) vertex Homer mesh, and an appropriate region \(R\) has been highlighted for each.
\cref{tab:denoising} presents the Shannon number, the number of non-zero wavelets, and the denoising results for each mesh for the same wavelet parameters as before --- \(\lambda=3\) and \(J_{0}=2\).
This straightforward hard-thresholding scheme boosts the signal-to-noise ratio of all signals on the meshes.
Note that the wavelet parameters \(\lambda{}\) and \(J_{0}\) have not been optimised for this denoising procedure.

\begin{table}
	\centering
	\caption{
		Denoising of other meshes.
	}\label{tab:denoising}
	\begin{tabular}{@{}rcccc@{}}
		\toprule
		        & Shannon       & Wavelets    & Init. \(\snr{z}\) & \(N_{\sigma}=2\ \snr{d}\) \\
		\midrule
		Cheetah & \(\num{72}\)  & \(\num{4}\) & \(\SI{0.32}{\dB}\)  & \(\SI{2.20}{\dB}\)        \\
		Dragon  & \(\num{169}\) & \(\num{5}\) & \(\SI{0.32}{\dB}\)  & \(\SI{1.44}{\dB}\)        \\
		Bird    & \(\num{194}\) & \(\num{5}\) & \(\SI{0.32}{\dB}\)  & \(\SI{1.41}{\dB}\)        \\
		Teapot  & \(\num{256}\) & \(\num{6}\) & \(\SI{0.32}{\dB}\)  & \(\SI{1.46}{\dB}\)        \\
		Cube    & \(\num{272}\) & \(\num{6}\) & \(\SI{0.32}{\dB}\)  & \(\SI{2.54}{\dB}\)        \\
		Homer   & \(\num{329}\) & \(\num{6}\) & \(\SI{0.32}{\dB}\)  & \(\SI{2.25}{\dB}\)        \\
		\bottomrule
	\end{tabular}
\end{table}

\section{Conclusions}\label{sec:conclusions}

Due to recent interest in geometric deep learning, many fields are extending existing methods in the Euclidean domain to the manifold or graph settings.
This work generalises Slepian scale-discretised wavelets on the sphere~\cite{Roddy2022} to Riemannian manifolds, whereby the wavelets themselves are restricted to a region of the manifold.
Slepian wavelets have many possible uses in various branches of science and engineering, where data only exists in a particular region.
A typical approach to problems of this kind is to analyse the signal with wavelets defined over the whole manifold, where spatially localised, scale-dependent features of the signal may be extracted.
Contamination of the wavelet coefficients at the boundaries of the region, however, still presents a problem.
Slepian scale-discretised wavelets (on manifolds) offer a potential solution to this obstacle.

The wavelets are built on the eigenfunctions of the Slepian spatial-spectral concentration problem (on the manifold), which are the basis functions of the region.
A tiling of the Slepian harmonic line allows one to construct scale-discretised wavelets, which: exhibit an explicit inversion formula, constitute a tight frame, and have excellent localisation properties in both spectral and spatial domains.
The wavelet transform is built on the sifting convolution~\cite{Roddy2021} which allows one to perform convolutions on (incomplete) manifolds --- which, in general, are not well-defined.
The wavelet construction here is analogous to the spherical setting, but where the basis functions are the eigenfunctions of the Laplace-Beltrami operator/eigenvectors of the graph Laplacian, rather than the spherical harmonics.

In computer graphics applications, three-dimensional shapes are often modelled by locally two-dimensional manifolds, which are then discretised as meshes.
A region is constructed on a mesh of Homer Simpson and the resulting Slepian functions are found.
A field is then placed on this mesh, and through a wavelet transform, the wavelets and corresponding wavelet coefficients of the field are presented.
The signal on the mesh is distorted with Gaussian white noise, and a straightforward denoising procedure is developed --- a common wavelet technique.
The wavelet coefficients are hard-thresholded with \(N_{\sigma}=2\), in which a boost in signal-to-noise ratio is observed.
This same denoising formalism is performed on some other meshes, resulting in a signal boost in all.
Slepian wavelets on manifolds can be used for many standard wavelet applications; for example, solving inverse problems via sparse regularisation~\cite{McEwen2013a,Wallis2017,Price2021}.

\bibliographystyle{IEEEtran}
\bibliography{library}

\end{document}